\documentclass[journal]{IEEEtran}

\ifCLASSINFOpdf
\else
\fi
\usepackage{cite}
\usepackage{amsmath,amssymb,amsfonts,amsthm}
\usepackage{algorithm}
\usepackage{algorithmic}
\usepackage{xcolor}
\usepackage{enumitem}
\usepackage{graphicx}
\usepackage{caption}
\usepackage{subcaption}
\usepackage{textcomp}
\usepackage[keeplastbox]{flushend}
\def\BibTeX{{\rm B\kern-.05em{\sc i\kern-.025em b}\kern-.08em
    T\kern-.1667em\lower.7ex\hbox{E}\kern-.125emX}}
\newcommand{\svdots}{\raisebox{3pt}{$\scalebox{.5}{\vdots}$}} 
\newcommand{\sddots}{\raisebox{3pt}{$\scalebox{.5}{$\ddots$}$}}

\newtheorem{theorem}{Theorem}

\newtheorem{corollary}{Corollary}

\begin{document}
%
\title{Performance of MIMO channel estimation with a physical model
\thanks{
This work has been performed in the framework of the Horizon 2020 project ONE5G (ICT-760809) receiving funds from the European Union. The authors would like to acknowledge the contributions of their colleagues in the project, although the views expressed in this contribution are those of the authors and do not necessarily represent the project.}
}
%
%
%
\author{Luc Le Magoarou, St\'ephane Paquelet
\thanks{Luc Le Magoarou and St\'ephane Paquelet are both with b$<>$com, Rennes, France. Contact addresses:  \texttt{luc.lemagoarou@b-com.com}, \texttt{stephane.paquelet@b-com.com}.}}
\markboth{Submitted}%
{Shell \MakeLowercase{\textit{et al.}}: Bare Demo of IEEEtran.cls for IEEE Journals}
%



\maketitle

\begin{abstract}
Channel estimation is challenging in multi-antenna communication systems, because of the large number of parameters to estimate. One way of facilitating this task is to use a physical model describing the multiple paths constituting the channel, in the hope of reducing the number of unknowns in the problem. The achievable performance of estimation using this kind of physical model is studied theoretically. It is found that adjusting the number of estimated paths leads to a bias-variance tradeoff which is characterized. Moreover, computing the Fisher information matrix of the model allows to identify orthogonal parameters, ultimately leading to fast and asymptotically optimal algorithms as a byproduct.  
\end{abstract}


\section{Introduction}

\IEEEPARstart{M}{ultiple}-input multiple-output (MIMO) communication systems allow for a dramatic increase in channel capacity, adding space to the classical time and frequency dimensions \cite{Telatar1999,Tse2005}. This is done by using several antennas at the transmitter ($N_t$) and at the receiver ($N_r$). The capacity of MIMO systems is maximized if the channel state is perfectly known at both ends of the link, which requires a channel estimation step.

Channel estimation is deeply impacted by the transition from single antenna to MIMO systems. Indeed, it amounts to determine a complex gain for each transmit/receive antenna pair, for each subcarrier. If $N_f$ subcarriers are used, the number of real parameters to estimate is thus $2N_rN_tN_f$, which may be very large for \emph{massive MIMO} systems, i.e.\ systems with up to several hundreds of antennas \cite{Rusek2013,Larsson2014}.

In a massive MIMO context, the classical least squares (LS) channel estimator is not adapted because the high dimensionality of the parameter space leads to an ill-posed problem. Therefore, in order to add prior information, it has been proposed classically to use Bayesian estimation and thus model the channel matrix as random with a known distribution, giving rise to estimators such as the linear minimum mean square error (LMMSE) \cite{Bjornson2017, Bazzi2017}. Another possibility is to use a parametric model based on the physics of wave propagation, in which the channel is expressed as a sum of $p$ paths \cite{Sayeed2002}. Whereas LS and LMMSE estimators have been studied extensively in terms of optimal training sequences and performance \cite{Biguesh2006}, a similar study is still lacking for channel estimators based on a physical model. 

\noindent {\bf Contributions and organization.} In this paper, the performance of MIMO wideband channel estimators based on a physical model is studied. The considered model is presented and the estimation error is decomposed into a bias and a variance terms in section~\ref{sec:problem_formulation}. Each term is then analyzed theoretically, beginning with the variance term in section~\ref{sec:variance} and the bias term in section~\ref{sec:bound_bias}, highlighting a bias-variance tradeoff piloted by the number of paths $p$ considered by the model. The theoretical analysis also allows to nicely interpret the channel estimation error and leads to the design of a computationally efficient channel estimation algorithm with optimality properties. Finally, the mathematical developments are assessed empirically in section~\ref{sec:experiments}, justifying the use of such physical models and showing the interest of the designed computationally efficient algorithm. 

Note that this paper is based on some of our previous work \cite{Lemagoarou2018,Lemagoarou2018b}, but significantly extends it in several ways. First, the wideband MIMO channel is considered here whereas only the narrowband MIMO channel was considered in \cite{Lemagoarou2018,Lemagoarou2018b}. Second, the assumptions required for the study of the variance are relaxed, and a much more extensive set of experiments is performed.


\section{Problem formulation}
\label{sec:problem_formulation}

\noindent{\bf Notations.} Matrices and vectors are denoted by bold upper-case and lower-case letters: $\mathbf{A}$ and $\mathbf{a}$ (except 3D ``\emph{spatial}'' vectors that are denoted $\overrightarrow{a}$); the $i$th column of a matrix $\mathbf{A}$ by $\mathbf{a}_i$; its entry at the $i$th line and $j$th column by $a_{ij}$. 
A matrix transpose, conjugate and transconjugate is denoted by $\mathbf{A}^T$, $\mathbf{A}^*$ and $\mathbf{A}^H$ respectively. The trace of a linear transformation represented by $\mathbf{A}$ is denoted $\text{Tr}(\mathbf{A})$.
For matrices $\mathbf{A}$ and $\mathbf{B}$, $\mathbf{A}\succeq \mathbf{B}$ means that $\mathbf{A}-\mathbf{B}$ is positive semidefinite.
 The linear span of a set of vectors $\mathcal{A}$ is denoted: $\text{span}(\mathcal{A})$. The Kronecker product and vectorization operators are denoted by $\otimes$ and $\text{vec}(\cdot)$ respectively, and the Hadamard (entry-wise) product by $\odot$. The identity matrix is denoted by $\mathbf{Id}$. $\mathcal{CN}(\boldsymbol\mu,\boldsymbol{\Sigma})$ denotes the standard complex gaussian distribution with mean $\boldsymbol\mu$ and covariance $\boldsymbol{\Sigma}$. $\mathbb{E}(\cdot)$ denotes expectation and $\text{cov}(\cdot)$ the covariance of its argument. $\mathcal{S}^2$ denotes the 3D unit sphere.

\subsection{Physical channel model}
\label{ssec:channel_model}
\begin{figure}[htbp]
\centering
\includegraphics[width=0.95\columnwidth]{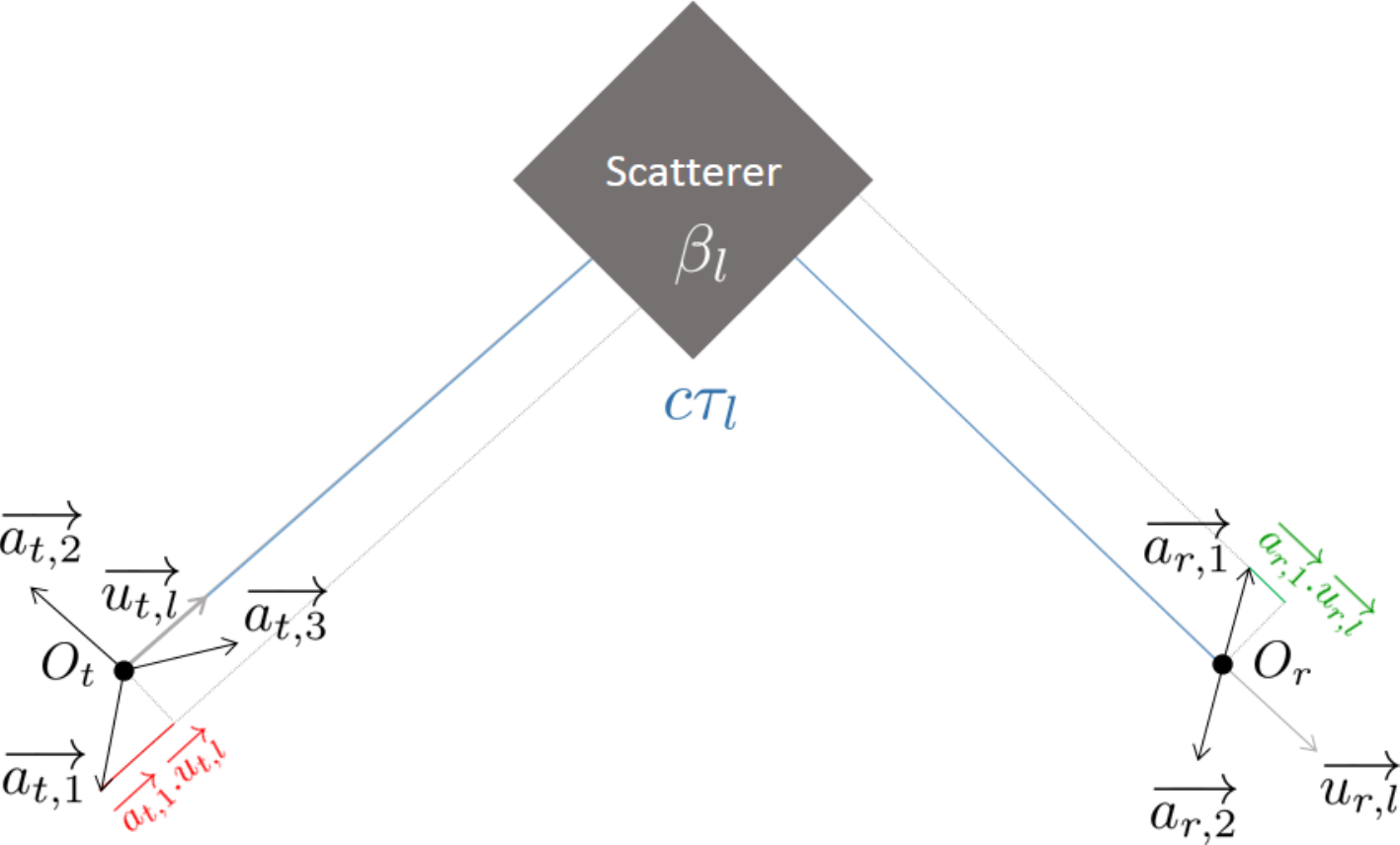}
\caption{Example of system representation for the $l$-th path. The length of the path from the first transmit antenna to the first receive antenna is equal to the length of the red line plus the length of the blue line minus the length of the green line, which illustrates \eqref{eq:channel}.}
\label{fig_system}
\end{figure}

Let us study a wideband block fading MIMO channel whose features are:
\begin{itemize}
\item $N_t$ transmit antennas located at the positions $\overrightarrow{a_{t,1}},\dots,\overrightarrow{a_{t,N_t}}$ with respect to the centroid of the transmit antenna array, whose radius is denoted $R_t \triangleq\text{max}_j \left\Vert\overrightarrow{a_{t,j}} \right\Vert_2 $.
\item $N_r$ receive antennas located at the positions $\overrightarrow{a_{r,1}},\dots,\overrightarrow{a_{r,N_r}}$ with respect to the centroid of the receive antenna array, whose radius is denoted $R_t \triangleq\text{max}_i \left\Vert\overrightarrow{a_{r,i}} \right\Vert_2 $.
\item $N_f$ subcarriers at the frequencies $f_c+f_1,\dots,f_c+f_{N_f}$, where $f_c$ is the central frequency ($\sum_k f_k=0$), and the bandwidth is denoted $B\triangleq f_{N_f} - f_1$.
\item $P$ propagation paths, where $\beta_l$ is the complex gain of the $l$-th path, $\tau_l$ is its delay, $\overrightarrow{u_{r,l}}$ is its direction of arrival (DoA) and $\overrightarrow{u_{t,l}}$ is its direction of departure (DoD).
\end{itemize}
Making the plane wave assumption, and further assuming $\frac{R_rB}{c} \ll 1$ and $\frac{R_tB}{c} \ll 1$ so that the phase difference due to the antenna positions is the same for all subcarriers, the channel between the $j$-th transmit antenna and the $i$-th receive antenna at the $k$-th subcarrier can be expressed
\begin{equation}
h_{ijk} = \sum_{l=1}^P \beta_l\mathrm{e}^{-\mathrm{j}2\pi\left[ \frac{1}{\lambda}\overrightarrow{a_{r,i}}.\overrightarrow{u_{r,l}}- \frac{1}{\lambda}\overrightarrow{a_{t,j}}.\overrightarrow{u_{t,l}} + f_k\tau_l \right]},
\label{eq:channel}
\end{equation}
where $\lambda \triangleq \frac{c}{f_c}$ is the wavelength at the central frequency. The used system representation is illustrated in figure~\ref{fig_system}.

It is possible to express the whole channel in a compact form by introducing the \emph{steering vectors},
$$
\mathbf{e}_t(\overrightarrow{u}) \triangleq \tfrac{1}{\sqrt{N_t}}\begin{pmatrix}
\mathrm{e}^{-\mathrm{j}\frac{2\pi}{\lambda} \overrightarrow{a_{t,1}}.\overrightarrow{u}},
\dots,
\mathrm{e}^{-\mathrm{j}\frac{2\pi}{\lambda} \overrightarrow{a_{t,N_t}}.\overrightarrow{u}}
\end{pmatrix}^T \in \mathbb{C}^{N_t},$$
$$
\mathbf{e}_r(\overrightarrow{u}) \triangleq \tfrac{1}{\sqrt{N_r}}\begin{pmatrix}
\mathrm{e}^{-\mathrm{j}\frac{2\pi}{\lambda} \overrightarrow{a_{r,1}}.\overrightarrow{u}},
\dots,
\mathrm{e}^{-\mathrm{j}\frac{2\pi}{\lambda} \overrightarrow{a_{r,N_r}}.\overrightarrow{u}}
\end{pmatrix}^T \in \mathbb{C}^{N_r},
$$
and the \emph{delay vector},
$$
\mathbf{e}_f(\tau) \triangleq \tfrac{1}{\sqrt{N_f}}\begin{pmatrix}
\mathrm{e}^{-\mathrm{j}2\pi f_1\tau},
\dots,
\mathrm{e}^{-\mathrm{j}2\pi f_{N_f}\tau}
\end{pmatrix}^T \in \mathbb{C}^{N_f}.
$$
Indeed, defining the \emph{characteristic vectors},  
$$
\mathbf{e}(\overrightarrow{u_r},\overrightarrow{u_t},\tau) \triangleq \mathbf{e}_f(\tau)
\otimes \mathbf{e}_t(\overrightarrow{u_t})^*\otimes
\mathbf{e}_r(\overrightarrow{u_r}) \in \mathbb{C}^{N_rN_tN_f},
$$
the channel can be expressed as a simple linear combination: 
\begin{equation}
\mathbf{h} = \sqrt{N_rN_tN_f}\sum_{l=1}^P\beta_l \mathbf{e}(\overrightarrow{u_{r,l}},\overrightarrow{u_{t,l}},\tau_l).
\label{eq:physical_channel}
\end{equation}
This expression of the channel is very general and encompasses many cases encountered in practice. For example, a narrowband (single carrier) channel corresponds to take $N_f = 1$ and $\mathbf{e}_f(\tau)=1,\, \forall \tau$, and a multiple-input single-output (MISO) channel corresponds to take $N_r = 1$ and $\mathbf{e}_r(\overrightarrow{u_r})=1,\, \forall \overrightarrow{u_r}$. Most MIMO channel simulators \cite{Sun2017,Jaeckel2014,3GPP2017} are based on
 a similar parameterization which can be described by a vector $\boldsymbol{\phi} \in \mathbb{R}^{N_{\boldsymbol{\phi}}}$ containing the \emph{physical parameters}:
\begin{equation}
 \boldsymbol{\phi} \triangleq (\beta_l,\overrightarrow{u_{r,l}},\overrightarrow{u_{t,l}},\tau_l)_{l=1}^P.
\label{eq:physical_params}
\end{equation}
The channel is thus described with $N_{\boldsymbol{\phi}} = 7P$ real numbers (two for the complex gain, two for each direction and one for the delay of each path), with $P$ taking values up to several hundreds in the most widepsread models \cite{Sun2017,Jaeckel2014,3GPP2017}.

\subsection{Channel estimation}
\label{ssec:channel_estimation}

\noindent{\bf Observations.}
Let us consider that in order to carry out channel estimation, $N_m$ noisy linear measurements of the channel are taken. The obtained observations are expressed
\begin{equation}
\mathbf{y} = \mathbf{Mh} + \mathbf{n},
\label{eq:observation}
\end{equation}
where $\mathbf{M} \in \mathbb{C}^{N_m\times N_rN_tN_f}$ is the \emph{observation matrix} and $\mathbf{n}$ is the received noise, which is assumed complex gaussian: $\mathbf{n}\sim \mathcal{CN}(\mathbf{0},\boldsymbol{\Sigma})$. Once again, this way of expressing the observations is very general. For example in the case of an hybrid system \cite{Elayach2014,Heath2016,Sayeed2016}, if the channel is observed through $N_c$ analog combiners given by the combining matrix $\mathbf{W} \in \mathbb{C}^{N_c\times N_r}$, and the same training sequence of duration $N_s$ given by the matrix $\mathbf{X} \in \mathbb{C}^{N_t\times N_s}$ is sent on $N_{ps}$ subcarriers whose positions are given by a row-sampled identity matrix $\mathbf{F}\in \{0,1\}^{N_{ps}\times N_f}$, then the observation matrix would take the form $\mathbf{M} = \mathbf{F}\otimes \mathbf{X}^T \otimes \mathbf{W}^H \in \mathbb{C}^{N_cN_sN_{ps} \times N_rN_tN_f}$. On the other hand, if every subcarrier is used to send pilots and the output of every receive antenna is observed, the observation matrix would take the form $\mathbf{M} = \mathbf{Id}_{N_f}\otimes \mathbf{X}^T \otimes \mathbf{Id}_{N_r}\in \mathbb{C}^{N_rN_sN_{f} \times N_rN_tN_f}$.

\noindent {\bf Objective.} Channel estimation aims at retrieving $\mathbf{h}$ from the observation of $\mathbf{y}$, knowing the observation matrix $\mathbf{M}$ and the distribution of the noise vector $\mathbf{n}$. The channel estimator is denoted $\hat{\mathbf{h}}$, and is assessed by its mean squared error
\begin{align}
\text{MSE}(\hat{\mathbf{h}}) &\triangleq \mathbb{E}\left[ \big\Vert \mathbf{h} - \hat{\mathbf{h}} \big\Vert_2^2 \right] \\
&=\big\Vert\mathbf{h}-\mathbb{E}[\hat{\mathbf{h}}]\big\Vert_2^2
+
\mathbb{E}\left[\big\Vert\hat{\mathbf{h}}-\mathbb{E}[\hat{\mathbf{h}}]\big\Vert_2^2\right],
\label{eq:MSE}
\end{align}
where the second line is the well-known bias-variance decomposition \cite{Kay1993}.
At first sight, the search space of channel estimation is of dimension $2N_rN_tN_f$ (number of real numbers needed to describe the channel vector), which may be very large in massive MIMO systems (up to several thousands). For this reason, classical estimation methods such as the least squares (LS) may not be appropriate. In order to overcome this limitation, some information about the channel has to be used to regularize the problem. For example, the channel can be considered as a random vector whose distribution is known, yielding Bayesian channel estimation \cite{Bjornson2017, Bazzi2017}. Another possibility is to use a physical channel model, as the one presented in section~\ref{ssec:channel_model}. In that case, according to \eqref{eq:physical_channel}, the dimension of the search space is equal to $7P$, which can also be very large. 

However in practice, the number of estimated paths $p$ is much smaller than $P$, so that there are only $N_{\boldsymbol{\theta}} = 7p$ parameters to estimate. In that case the estimate belongs by construction to the set $\mathcal{M}_p$ of vectors that can be expressed by a sum of at most $p$ \emph{virtual paths}.  This set is called \emph{model set} hereafter and is expressed 
 $$\mathcal{M}_p=\big\{ \mathbf{a} \in \mathbb{C}^{N_r N_t N_f}|\, \mathbf{a} = \sum\nolimits_{n=1}^p  \gamma_n\mathbf{e}(\overrightarrow{w_{r,n}},\overrightarrow{w_{t,n}},\xi_n)\big\},$$
with $\alpha_n \in \mathbb{C}$, $\overrightarrow{w_{r,n}} \in \mathcal{S}^2$, $\overrightarrow{w_{t,n}} \in \mathcal{S}^2$ and $\xi_n \in \mathbb{R}_+$. Such sets obey the inclusion relation $\mathcal{M}_{q} \subset \mathcal{M}_{q+1}$. 
The best element of the model set is denoted
\begin{equation}
\mathbf{h}_{\mathcal{M}_p} \triangleq \underset{\mathbf{g}\in\mathcal{M}_p}{\text{argmin}}\left\Vert \mathbf{h} - \mathbf{g} \right\Vert_2.
\end{equation}
It is simply the projection of the channel onto the model set $\mathcal{M}_p$. By construction, this projection can be written 
 \begin{equation}
\mathbf{h}_{\mathcal{M}_p} = \sqrt{N_rN_tN_f}\sum_{n=1}^p\gamma_n \mathbf{e}(\overrightarrow{w_{r,n}},\overrightarrow{w_{t,n}},\xi_n),
\label{eq:projected_channel}
\end{equation}
which gives optimal values for the estimated parameters, or \emph{virtual parameters}
 \begin{equation}
 \boldsymbol{\theta} \triangleq (\gamma_n,\overrightarrow{w_{r,n}},\overrightarrow{w_{t,n}},\xi_n)_{n=1}^p \in \mathbb{R}^{N_{\boldsymbol{\theta}}}.
 \label{eq:true_parameters}
 \end{equation}
 The actual estimate will be denoted 
 \begin{equation}
\hat{\mathbf{h}} = \sqrt{N_rN_tN_f}\sum_{m=1}^p\alpha_m \mathbf{e}(\overrightarrow{v_{r,m}},\overrightarrow{v_{t,m}},\zeta_m),
\label{eq:estimator_channel}
\end{equation}
so that the estimated parameters are
 \begin{equation}
 \hat{\boldsymbol{\theta}} \triangleq (\alpha_m,\overrightarrow{v_{r,m}},\overrightarrow{v_{t,m}},\zeta_m)_{m=1}^p \in \mathbb{R}^{N_{\boldsymbol{\theta}}}.
 \label{eq:estimated_parameters}
 \end{equation}
 In summary, the channel depends on the $7P$ physical parameters $\boldsymbol{\phi}$ given in \eqref{eq:physical_params}, but the considered model depends on only $7p$ parameters whose optimal values make up the virtual parameters $\boldsymbol{\theta}$ defined in \eqref{eq:true_parameters}, and the channel estimate depends on the actually estimated parameters $\hat{\boldsymbol{\theta}}$ given in \eqref{eq:estimated_parameters}. The central question of this paper is:\begin{center}
 \emph{How well can an estimator that yields estimates belonging to $\mathcal{M}_p$ perform with a channel taking the form \eqref{eq:physical_channel}?}
 \end{center}

 
\noindent{\bf MSE decomposition.}
In order to answer this fundamental question, let us first notice that an estimate taking the form \eqref{eq:estimator_channel} cannot be better than the best element of the model set, that is
$$
\big\Vert \mathbf{h} - \hat{\mathbf{h}} \big\Vert_2 \geq \left\Vert \mathbf{h} - \mathbf{h}_{\mathcal{M}_p} \right\Vert_2.
$$
The study of this paper focuses on estimators whose expectation $\mathbb{E}[\hat{\mathbf{h}}]$ is equal to the projection $\mathbf{h}_{\mathcal{M}_p}$, so that the MSE can be decomposed as in \eqref{eq:MSE}:
\begin{equation}
\text{MSE}(\hat{\mathbf{h}}) = 
\big\Vert\mathbf{h}-\mathbf{h}_{\mathcal{M}_p}\big\Vert_2^2
+
\mathbb{E}\left[\big\Vert\hat{\mathbf{h}}-\mathbf{h}_{\mathcal{M}_p}\big\Vert_2^2\right].
\label{eq:bias_variance}
\end{equation}
According to this assumption, the bias of the estimator is identified to the \emph{model bias} (distance of the true channel $\mathbf{h}$ to the model $\mathcal{M}_p$), and the variance is computed with respect to the projection $\mathbf{h}_{\mathcal{M}_p}$. In the next sections, the two terms of this decomposition are analyzed separately.

\section{Variance analysis} 
\label{sec:variance}

In this section, the variance term of the mean squared error \eqref{eq:bias_variance} is bounded using the Cram\'er-Rao Bound (CRB) \cite{Rao1945,Cramer1946}, which is valid for any unbiased estimator. The case of a linear channel parameterization is first evoked in section~\ref{ssec:linear} because of its remarkable similarities with the studied problem. The bound for the studied model is then derived in section~\ref{ssec:cramer_rao_bound}, and computationally efficient estimation algorithms are deduced of the particular form of the Fisher information matrix in sections~\ref{ssec:fim} and~\ref{ssec:algos}.

\subsection{The linear case}
\label{ssec:linear}
The projected channel $\mathbf{h}_{\mathcal{M}_p}$ is a nonlinear function of the virtual parameters $\boldsymbol{\theta}$. Let us first look at a simplified problem in which the channel to estimate $\mathbf{h}_l$ is linearly linked to the parameter vector $\boldsymbol{\theta}_l \in \mathbb{R}^{N_{\boldsymbol{\theta}}} $ via the equation $\mathbf{h}_l \triangleq \mathbf{D}\boldsymbol{\theta}_l $. In that hypothetical case, the observations would read
$$
\mathbf{y}_l = \mathbf{MD}\boldsymbol{\theta}_l + \mathbf{n}.
$$
The least-squares estimate of the parameter vector (without constraining it to be real) is then
\begin{align*}
\hat{\boldsymbol{\theta}}_l &= (\mathbf{D}^H\mathbf{M}^H\mathbf{MD})^{-1}\mathbf{D}^H\mathbf{M}^H\mathbf{y}_l \\
&= \boldsymbol{\theta}_l + (\mathbf{D}^H\mathbf{M}^H\mathbf{MD})^{-1}\mathbf{D}^H\mathbf{M}^H\mathbf{n},
\end{align*}
and a channel estimate can be obtained as
\begin{align*}
\hat{\mathbf{h}}_l &= \mathbf{D}\hat{\boldsymbol{\theta}}_l \\
&= \underbrace{\mathbf{D}\boldsymbol{\theta}_l}_{\mathbf{h}_l} + \underbrace{\mathbf{D}(\mathbf{D}^H\mathbf{M}^H\mathbf{MD})^{-1}\mathbf{D}^H\mathbf{M}^H\mathbf{n}}_{\mathbf{n}'}.
\end{align*}
This is an unbiased estimator of $\mathbf{h}_l$. Assuming a white noise $\mathbf{n}\sim \mathcal{CN}(\mathbf{0},\sigma^2\mathbf{Id})$, the covariance of $\mathbf{n}'$ is given by
$$
\mathbb{E}[\mathbf{n}'\mathbf{n}'^H] = \sigma^2\mathbf{D}(\mathbf{D}^H\mathbf{M}^H\mathbf{MD})^{-1}\mathbf{D}^H.
$$
The variance of the estimator $\hat{\mathbf{h}}_l$ is thus given by
\begin{align*}
\mathbb{E}\left[\big\Vert\hat{\mathbf{h}}_l-\mathbf{h}_{l}\big\Vert_2^2\right] &= \sigma^2\text{Tr}[\mathbf{D}(\mathbf{D}^H\mathbf{M}^H\mathbf{MD})^{-1}\mathbf{D}^H]
\\
&= \frac{\sigma^2N_{\boldsymbol{\theta}}}{\Vert \mathbf{M} \Vert_2^2} \;\text{if}\; \mathbf{D}^H\frac{\mathbf{M}^H\mathbf{M}}{\Vert \mathbf{M} \Vert_2^2}\mathbf{D} = \mathbf{D}^H\mathbf{D}.
\end{align*}
According to the second line of the above equation, provided the observation matrix fulfills some condition (being conformal on the column space of $\mathbf{D}$), the variance is proportional to the number of parameters and the noise variance, and inversely proportional to the observation matrix squared norm.
As will be highlighted in the next subsection, the lower bound on the variance of $\hat{\mathbf{h}}$ is surprisingly similar to this expression, despite the non-linearities involved.

\subsection{Cram\'er-Rao bound}
\label{ssec:cramer_rao_bound}

 Let us go back to the original problem and consider the virtual parameters $\boldsymbol{\theta}$ defined in \eqref{eq:true_parameters} as true parameters and $\hat{\boldsymbol{\theta}}$ defined in \eqref{eq:estimated_parameters} as their unbiased estimates.
Applying the complex CRB \cite{Vandenbos1994} yields 
$$
\mathbb{E}\left[\big\Vert\hat{\mathbf{h}}-\mathbf{h}_{\mathcal{M}_p}\big\Vert_2^2\right] \geq \text{Tr}\left[\frac{\partial \mathbf{h}_{\mathcal{M}_p}}{\partial \boldsymbol{\theta}}\mathbf{I}(\boldsymbol{\theta})^{-1}\frac{\partial \mathbf{h}_{\mathcal{M}_p}}{\partial \boldsymbol{\theta}}^H\right] \triangleq \text{CRB},
$$
with $
 \frac{\partial \mathbf{h}_{\mathcal{M}_p}}{\partial \boldsymbol{\theta}}
\triangleq
\big(\frac{\partial \mathbf{h}_{\mathcal{M}_p}}{\partial \theta_1},\dots,\frac{\partial \mathbf{h}_{\mathcal{M}_p}}{\partial \theta_{N_{\boldsymbol{\theta}}}} \big) \in \mathbb{C}^{N_rN_tN_f\times N_{\boldsymbol{\theta}}},
$ and $\mathbf{I}(\boldsymbol{\theta}) \in \mathbb{R}^{ N_{\boldsymbol{\theta}}\times N_{\boldsymbol{\theta}}}$ being the Fisher information matrix (FIM).
Noticing that the observation $\mathbf{y}$ follows a complex gaussian distribution,
$$
\mathbf{y} \sim \mathcal{CN}\big(\underbrace{\mathbf{M} \mathbf{h}}_{\boldsymbol{\mu}(\boldsymbol{\theta})},\boldsymbol{\Sigma} \big), $$ the FIM is given by the Slepian-Bangs formula \cite{Slepian1954,Bangs1971}: 
\begin{align}
\mathbf{I}(\boldsymbol{\theta}) &= 2\mathfrak{Re}\left\{\frac{\partial \boldsymbol{\mu}(\boldsymbol{\theta})}{\partial \boldsymbol{\theta}}^H\boldsymbol{\Sigma}^{-1}\frac{\partial \boldsymbol{\mu}(\boldsymbol{\theta})}{\partial \boldsymbol{\theta}}\right\}
\\
&= 2\mathfrak{Re}\left\{\frac{\partial \mathbf{h}_{\mathcal{M}_p}}{\partial \boldsymbol{\theta}}^H\mathbf{M}^H\boldsymbol{\Sigma}^{-1}\mathbf{M}\frac{\partial \mathbf{h}_{\mathcal{M}_p}}{\partial \boldsymbol{\theta}}\right\}, 
\label{eq:FIM_gen}
\end{align}
 where the second line is true assuming that locally, $\mathbf{h}$ can be decomposed as $\mathbf{h} = \mathbf{h}_{\mathcal{M}_p} + \mathbf{r}$, with $\mathbf{r}$ not being a function of the virtual parameters $\boldsymbol{\theta}$, so that $\frac{\partial \mathbf{h}}{\partial \boldsymbol{\theta}} = \frac{\partial \mathbf{h}_{\mathcal{M}_p}}{\partial \boldsymbol{\theta}}$.
 Let us now state the main result of the paper regarding the variance.
 \begin{theorem}
 Provided the model is identifiable,
\begin{equation}
\text{CRB}\geq \frac{N_{\boldsymbol{\theta}}}{2\left\Vert \mathbf{M}^H\boldsymbol{\Sigma}^{-1}\mathbf{M} \right\Vert_2},
\label{eq:thm_variance}
\end{equation}
with equality if the condition $$C_{\text{opt}} : \frac{\partial \mathbf{h}_{\mathcal{M}_p}^H}{\partial \boldsymbol{\theta}}\frac{\mathbf{M}^H\boldsymbol{\Sigma}^{-1}\mathbf{M}}{\Vert\mathbf{M}^H\boldsymbol{\Sigma}^{-1}\mathbf{M}\Vert_2}\frac{\partial \mathbf{h}_{\mathcal{M}_p}}{\partial \boldsymbol{\theta}} = \frac{\partial \mathbf{h}_{\mathcal{M}_p}^H}{\partial \boldsymbol{\theta}}\frac{\partial \mathbf{h}_{\mathcal{M}_p}}{\partial \boldsymbol{\theta}}$$ is fulfilled.
\label{thm:variance}
\end{theorem}
The theorem is proven in appendix~\ref{app:proof_thm_variance}, in order to keep the flow of the paper.

An important feature of this result is that the optimal CRB is proportional to the number of parameters, and consequently to the number of estimated paths $p$, if $C_{\text{opt}}$ is fulfilled. However, the condition $C_{\text{opt}}$ may seem a bit abstract in its current general form. In order to ease interpretation, let us consider a special case given by the following corollary.

\begin{corollary}
For a white gaussian noise with $\boldsymbol{\Sigma} = \sigma^2\mathbf{Id}$, the bound of theorem~\ref{thm:variance} becomes 
\begin{equation}
\text{CRB}\geq \frac{N_{\boldsymbol{\theta}}\sigma^2}{2\left\Vert \mathbf{M} \right\Vert_2^2},
\end{equation}
with equality if $C_{\text{opt}}':\frac{\partial \mathbf{h}_{\mathcal{M}_p}^H}{\partial \boldsymbol{\theta}}\frac{\mathbf{M}^H\mathbf{M}}{\left\Vert\mathbf{M}\right\Vert_2^2}\frac{\partial \mathbf{h}_{\mathcal{M}_p}}{\partial \boldsymbol{\theta}} = \frac{\partial \mathbf{h}_{\mathcal{M}_p}^H}{\partial \boldsymbol{\theta}}\frac{\partial \mathbf{h}_{\mathcal{M}_p}}{\partial \boldsymbol{\theta}}$.
\label{cor:variance}
\end{corollary}
Note that this bound is very similar to the variance obtained in the linear case in section \ref{ssec:linear}, except for the division by two (due to the fact that the parameters being real was overlooked in section \ref{ssec:linear}). This means that, at least asymptotically, the nonlinear model behaves like a linear one (provided an efficient estimator is available in the nonlinear case).

{\noindent\bf Interpretations.} The condition $C_{\text{opt}}'$ is quite easily understood. It means that the observation matrix $\mathbf{M}$ has to preserve angles (be conformal) on the column space of $ \frac{\partial \mathbf{h}_{\mathcal{M}_p}}{\partial \boldsymbol{\theta}}$. One obvious although conservative way to fulfill $C_{\text{opt}}'$ is then to take $\mathbf{M}^H\mathbf{M} = \alpha^2\mathbf{Id}$, so that $\mathbf{M}$ is conformal on the whole space $\mathbb{C}^{N_rN_tN_f}$ (including the column space of $ \frac{\partial \mathbf{h}_{\mathcal{M}_p}}{\partial \boldsymbol{\theta}}$). This is possible only if $N_m\geq N_rN_tN_f$, meaning that the number of measurements has to be greater than the dimension of the channel. Another (more clever) way of fulfilling $C_{\text{opt}}'$ is to take $\mathbf{M} = \alpha\mathbf{QU}^H$ where $\mathbf{Q} \in \mathbb{C}^{N_{\boldsymbol{\theta}} \times N_{\boldsymbol{\theta}}}$ is an unitary matrix and $\mathbf{U} \in \mathbb{C}^{N_rN_tN_f \times N_{\boldsymbol{\theta}}}$ has its columns forming an orthogonal basis of the column space of $ \frac{\partial \mathbf{h}_{\mathcal{M}_p}}{\partial \boldsymbol{\theta}}$. That way, $N_m = N_{\boldsymbol{\theta}}$, which may be much smaller than the channel dimension $N_rN_tN_f$.

Let us now look at the quantity $\left\Vert \mathbf{M} \right\Vert_2^2$ when using the two aforementioned strategies with a fixed power per measurement $P_m$. First, if each coordinate of $\mathbb{C}^{N_rN_tN_f}$ is measured $K$ times, this yields $\mathbf{M} = \sqrt{P_m} (\mathbf{Id}^{(1)},\dots, \mathbf{Id}^{(K)})^T$ and $N_m = K N_rN_tN_f$. Second, if each basis vector of the column space of $ \frac{\partial \mathbf{h}_{\mathcal{M}_p}}{\partial \boldsymbol{\theta}}$ is measured $K$ times, this yields $\mathbf{M} = \sqrt{P_m} ({\mathbf{UQ}^H}^{(1)},\dots, {\mathbf{UQ}^H}^{(K)})^H$ and $N_m = K N_{\boldsymbol{\theta}}$. In these two cases, $\left\Vert \mathbf{M} \right\Vert_2^2 = P_mK$ so that $\text{CRB} = \frac{N_{\boldsymbol{\theta}}\sigma^2}{2P_mK}$. The bound is thus in both cases inversely proportional to the measurements power, as well as to the number of measurements taken. However the second strategy is much more computationally efficient since it requires to take only $N_m = K N_{\boldsymbol{\theta}}$ measurements instead of $N_m = K N_rN_tN_f$. Note that $P_m$ is strongly linked to the transmit power, for example in the case of a single subcarrier and an observation matrix of the form $\mathbf{M} = \mathbf{X}^T \otimes \mathbf{Id}_{N_r}$ where each column of $\mathbf{X}$ has its squared norm equal to $P_t$, then $P_m=P_t$. 

A study of practical and resource efficient pilot designs that fulfill (exactly or approximately) $C_{\text{opt}}$ would be of great interest, but is beyond the scope of this paper. The condition $C_{\text{opt}}$ is assumed fulfilled in the remaining of this paper, yielding an \emph{optimal observation matrix}.

\subsection{Fisher information matrix}
\label{ssec:fim}

The CRB computed in the previous subsection is attained by efficient estimators \cite{Kay1993}. Maximum likelihood estimators (MLEs) are asymptotically efficient, and The Fisher information matrix (FIM) determines their asymptotic properties (it is the expected Hessian of the negative log-likelihood), and can be used to design efficient estimation algorithms. Let us compute the FIM for the considered physical model \eqref{eq:true_parameters}. 
It exhibits a block structure
$$
\mathbf{I}(\boldsymbol{\theta}) =
\left(\begin{smallmatrix}
\mathbf{I}^{(1,1)} & \mathbf{I}^{(1,2)}&\dots& \mathbf{I}^{(1,p)} \\
\mathbf{I}^{(2,1)} & \mathbf{I}^{(2,2)}&& \\
\svdots&&\sddots&\\
\mathbf{I}^{(p,1)} & && \mathbf{I}^{(p,p)}
\end{smallmatrix}\right),
$$
where, according to \eqref{eq:FIM_gen} assuming $C_\text{opt}$, the off-diagonal block $\mathbf{I}^{(m,n)} \in \mathbb{R}^{7\times 7}$ contains the correlations between the sensitivities of the channel to parameters of the $m$-th and $n$-th paths, and the diagonal block $\mathbf{I}^{(n,n)}\in \mathbb{R}^{7\times 7}$ contains the correlations between the sensitivities to parameters of the $n$-th path. Let us focus here on the diagonal blocks, since as will be stated in more details in the next subsection, most estimation algorithms do not handle paths jointly but sequentially, implicitly assuming distinct estimated paths have uncorrelated effects on the channel. 

It remains to compute the derivatives (sensitivities) $\frac{\partial \mathbf{h}_{\mathcal{M}_p}}{\partial \boldsymbol{\theta}}$ with $\boldsymbol{\theta}$ defined in \eqref{eq:true_parameters}. Writing $\gamma_n = \rho_n\mathrm{e}^{\mathrm{j}\phi_n}$ yields
$$
\frac{\partial \mathbf{h}_{\mathcal{M}_p}}{\partial \rho_n} = \sqrt{N_rN_tN_f}\mathrm{e}^{\mathrm{j}\phi_n} \mathbf{e}_f(\xi_n)
\otimes \mathbf{e}_t(\overrightarrow{w_{t,n}})^*\otimes
\mathbf{e}_r(\overrightarrow{w_{r,n}})
$$
and
$$
\frac{\partial \mathbf{h}_{\mathcal{M}_p}}{\partial \phi_n} = \sqrt{N_rN_tN_f}\mathrm{j}\gamma_n \mathbf{e}_f(\xi_n)
\otimes \mathbf{e}_t(\overrightarrow{w_{t,n}})^*\otimes
\mathbf{e}_r(\overrightarrow{w_{r,n}}).
$$
Then, denoting $\overrightarrow{b_{r,n,1}},\overrightarrow{b_{r,n,2}}$ the two basis vectors used to describe a change in $\overrightarrow{w_{r,n}}$ and $\overrightarrow{b_{t,n,1}},\overrightarrow{b_{t,n,2}}$ those used to describe a change in $\overrightarrow{w_{t,n}}$,
$$
\frac{\partial \mathbf{h}_{\mathcal{M}_p}}{\partial \overrightarrow{b_{r,n,i}}} = \sqrt{N_rN_tN_f}\gamma_n \mathbf{e}_f(\xi_n)
\otimes \mathbf{e}_t(\overrightarrow{w_{t,n}})^*\otimes(
\mathbf{a}_{r,n,i}\odot\mathbf{e}_r(\overrightarrow{w_{r,n}}))
$$
with
$
\mathbf{a}_{r,n,i}\triangleq -\mathrm{j}\frac{2\pi}{\lambda} 
(
\overrightarrow{a_{r,1}}.\overrightarrow{b_{r,n,i}},\dots,
\overrightarrow{a_{r,N_r}}.\overrightarrow{b_{r,n,i}}
)^T$, and $$
\frac{\partial \mathbf{h}_{\mathcal{M}_p}}{\partial \overrightarrow{b_{t,n,i}}} = \sqrt{N_rN_tN_f}\gamma_n \mathbf{e}_f(\xi_n)
\otimes (\mathbf{a}_{t,n,i}\odot\mathbf{e}_t(\overrightarrow{w_{t,n}}))^*\otimes\mathbf{e}_r(\overrightarrow{w_{r,n}})
$$
with
 $
\mathbf{a}_{t,n,i}\triangleq -\mathrm{j}\frac{2\pi}{\lambda} 
(
\overrightarrow{a_{r,1}}.\overrightarrow{b_{t,n,i}},
\dots,
\overrightarrow{a_{r,N_t}}.\overrightarrow{b_{t,n,i}}
)^T
$.
Finally,
$$
\frac{\partial \mathbf{h}_{\mathcal{M}_p}}{\partial \xi_n} = \sqrt{N_rN_tN_f}\gamma_n (\mathbf{f}\odot\mathbf{e}_f(\xi_n))
\otimes \mathbf{e}_t(\overrightarrow{w_{t,n}})^*\otimes
\mathbf{e}_r(\overrightarrow{w_{r,n}})
$$
with
$
\mathbf{f} \triangleq -\mathrm{j}2\pi 
(f_1,\dots,f_{N_f})^T.
$

Note that in section~\ref{ssec:channel_model} the antenna locations are expressed with respect to the centroid and the frequencies with respect to the central frequency. This is not an arbitrary choice.
Indeed, this carefully chosen parameterization guarantees that 
$$\mathbf{1}^T\mathbf{f} = \mathbf{1}^T\mathbf{a}_{r,n,i} = \mathbf{1}^T\mathbf{a}_{t,n,i} = 0,$$ 
where $\mathbf{1}$ is a vector of the appropriate size having all entries equal to one. Taking this into account and injecting these derivatives in \eqref{eq:FIM_gen} assuming $C_{\text{opt}}$ yields
\begin{equation}
\mathbf{I}^{(n,n)} = C
\left(\begin{array}{ccccc}
1
&
0
&
\mathbf{0}
&
\mathbf{0}
&
0
\\ \\
0
&
\rho_n^2
&
\mathbf{0}
&
\mathbf{0}
&
0
\\ \\
\mathbf{0}
&
\mathbf{0}
&
\mathbf{B}_{r,n}
&
\mathbf{0}
&
\mathbf{0}
\\ \\
\mathbf{0}
&
\mathbf{0}
&
\mathbf{0}
&
\mathbf{B}_{t,n}
&
\mathbf{0}
\\ \\
0
&
0
&
\mathbf{0}
&
\mathbf{0}
&
\rho_n^2\left\Vert \mathbf{f} \right\Vert_2^2
\end{array}\right)
\label{eq:FIM}
\end{equation}
with $C = N_rN_tN_f\left\Vert \mathbf{M}^H\boldsymbol{\Sigma}^{-1}\mathbf{M} \right\Vert_2$, $\mathbf{0}$ denotes zero vectors or matrices of appropriate size,
$$
\mathbf{B}_{r,n} \triangleq \rho_n^2\begin{pmatrix}
\left\Vert \mathbf{a}_{r,n,1} \right\Vert_2^2 & \mathbf{a}_{r,n,1}^H\mathbf{a}_{r,n,2}  \\
\mathbf{a}_{r,n,2}^H\mathbf{a}_{r,n,1}& \left\Vert \mathbf{a}_{r,n,2} \right\Vert_2^2
\end{pmatrix}
$$
and 
$$
\mathbf{B}_{t,n} \triangleq \rho_n^2\begin{pmatrix}
\left\Vert \mathbf{a}_{t,n,1} \right\Vert_2^2 & \mathbf{a}_{t,n,1}^H\mathbf{a}_{t,n,2}  \\
\mathbf{a}_{t,n,2}^H\mathbf{a}_{t,n,1}& \left\Vert \mathbf{a}_{t,n,2} \right\Vert_2^2
\end{pmatrix}.
$$
The important feature of \eqref{eq:FIM} is that the FIM $\mathbf{I}^{(n,n)}$ being block diagonal, the parameters of the same path are orthogonal to each other (thanks to the chosen parameterization). Parameter orthogonality has several implications \cite{Cox1987}. It can for instance be exploited to design efficient estimation algorithms, as is the topic of the next subsection. The diagonal blocks of the FIM are not analyzed here, since they do not have any impact on the CRB as long as the model is identifiable (some interpretations regarding the diagonal terms in a simplified setting are available in \cite{Lemagoarou2018}).

\subsection{Efficient estimation algorithms}
\label{ssec:algos}
Let us focus on algorithms aimed at obtaining an estimate taking the form of \eqref{eq:estimator_channel}. First, one can rewrite \eqref{eq:estimator_channel} as 
$$
\hat{\mathbf{h}} = \mathbf{E}\boldsymbol{\alpha},
$$
with $\mathbf{E} \triangleq \left(\mathbf{e}(\overrightarrow{v_{r,1}},\overrightarrow{v_{t,1}},\zeta_1),\dots, \mathbf{e}(\overrightarrow{v_{r,p}},\overrightarrow{v_{t,p}},\zeta_p)\right)$ and $\boldsymbol{\alpha} \triangleq \sqrt{N_rN_tN_f}(\alpha_1,\dots,\alpha_p)^T$.
Assuming a white gaussian noise, maximum likelihood channel estimation corresponds to solve the optimization problem
\begin{equation}
    \label{eq:ML2}
\underset{\mathbf{E},\boldsymbol{\alpha}}{\text{minimize}}\;\big\Vert \mathbf{y} -  \mathbf{M}\mathbf{E}\boldsymbol{\alpha}\big\Vert_2^2,\quad \hat{\mathbf{h}} \leftarrow \mathbf{E}\boldsymbol{\alpha}.
\end{equation}
Note that given $\mathbf{E}$, the optimal vector $\boldsymbol{\alpha}$ can be obtained as the solution of a least squares problem as $\boldsymbol{\alpha}_{\text{opt}} = (\mathbf{E}^H\mathbf{M}^H\mathbf{ME})^{-1}\mathbf{E}^H\mathbf{M}^H\mathbf{y}$, so that in the end channel estimation amounts to find an optimal $\mathbf{E}$, i.e.\ an optimal set of $p$ vectors $\left\{\mathbf{e}(\overrightarrow{v_{r,1}},\overrightarrow{v_{t,1}},\zeta_1),\dots, \mathbf{e}(\overrightarrow{v_{r,p}},\overrightarrow{v_{t,p}},\zeta_p)\right\}$.

\noindent \textbf{Greedy strategy.} Looking for the $p$ vectors jointly yields a very complex optimization problem. Instead, greedy strategies have been proposed which consist in building a dictionary of characteristic vectors and applying a sparse recovery algorithm such as orthogonal matching pursuit (OMP) \cite{Mallat1993,Tropp2007,Bajwa2010}. This amounts to estimate the paths one by one, i.e.\ building the matrix $\mathbf{E}$ column by column. Denoting $\mathbf{E}^{(k)} \triangleq \left(\mathbf{e}(\overrightarrow{v_{r,1}},\overrightarrow{v_{t,1}},\zeta_1),\dots, \mathbf{e}(\overrightarrow{v_{r,k}},\overrightarrow{v_{t,k}},\zeta_k)\right)$ the state of the matrix $\mathbf{E}$ at the $k$-th iteration, the optimal vector $\boldsymbol{\alpha}^{(k)} \gets (\mathbf{E}^{(k)H}\mathbf{M}^H\mathbf{M}\mathbf{E}^{(k)})^{-1}\mathbf{E}^{(k)H}\mathbf{M}^H\mathbf{y}$ is computed so that a residual $\mathbf{r}^{(k+1)} \gets \mathbf{y} - \mathbf{ME}^{(k)} \boldsymbol{\alpha}^{(k)}$ is used at the next iteration. Such a strategy is summarized in algorithm~\ref{algo_greedy}.

%
%
%
%
%

\begin{algorithm}[htb]
\caption{Greedy channel estimation}
\begin{algorithmic}[1] 
\REQUIRE Observation $\mathbf{y}$, observation matrix $\mathbf{M}$, number of paths to estimate $p$.
\STATE $\mathbf{r}^{(1)}\leftarrow \mathbf{y}$
\FOR{$i=1,\dots,p$}
\STATE Estimate a characteristic vector $\mathbf{e}(\overrightarrow{v_{r,i}},\overrightarrow{v_{t,i}},\zeta_i)$ based on $\mathbf{r}^{(i)}$
\STATE $\mathbf{E}^{(i)} \leftarrow \left(\mathbf{e}(\overrightarrow{v_{r,1}},\overrightarrow{v_{t,1}},\zeta_1),\dots, \mathbf{e}(\overrightarrow{v_{r,i}},\overrightarrow{v_{t,i}},\zeta_i)\right)$
\STATE Update the coefficients: \\
$\boldsymbol{\alpha}^{(i)} \leftarrow \left(\mathbf{E}^{(i) H}\mathbf{M}^H\mathbf{M}\mathbf{E}^{(i)}\right)^{-1}\mathbf{E}^{(i) H}\mathbf{M}^H\mathbf{y}$
\STATE Update the residual: \\ 
$\mathbf{r}^{(i+1)} \leftarrow \mathbf{y} - \mathbf{M}\mathbf{E}^{(i)}\boldsymbol{\alpha}^{(i)}$
\ENDFOR
\STATE $\hat{\mathbf{h}} \leftarrow \mathbf{E}^{(p)}\boldsymbol{\alpha}^{(p)}$
\ENSURE Channel estimate $\hat{\mathbf{h}}$
\end{algorithmic}
\label{algo_greedy}
\end{algorithm}

The critical and most computationally intensive part of the above algorithm is the characteristic vector estimation (line 3), which in the maximum likelihood framework amounts to solve the optimization problem
$$\underset{\mathbf{x}}{\text{maximize}} \;f(\mathbf{x},\mathbf{r}^{(i)})\triangleq \frac{|\mathbf{x}^H\mathbf{M}^H\mathbf{r}^{(i)}|^2}{\left\Vert \mathbf{Mx} \right\Vert_2^2},$$
with $\mathbf{x}$ being a characteristic vector.

\noindent {\bf Joint estimation.} One straightforward and popular way to solve this problem is given in algorithm~\ref{algo_joint} which, if testing $N_\zeta$ delays, $N_{\overrightarrow{v_r}}$ DoAs and $N_{\overrightarrow{v_t}}$ DoDs requires $N_\zeta N_{\overrightarrow{v_r}} N_{\overrightarrow{v_t}}$ computations of the cost function $f(\mathbf{x},\mathbf{r}^{(i)})$.
\begin{algorithm}[htb]
\caption{Joint characteristic vector estimation}
\begin{algorithmic}[1] 
\REQUIRE Current residual $\mathbf{r}^{(i)}$, observation matrix $\mathbf{M}$
\STATE Estimate physical parameters $\overrightarrow{v_{r,i}},\overrightarrow{v_{t,i}},\zeta_i$ based on $\mathbf{r}^{(i)}$ \\
$\overrightarrow{v_{r,i}},\overrightarrow{v_{t,i}},\zeta_i\leftarrow \underset{\overrightarrow{v_{r}},\overrightarrow{v_{t}},\zeta}{\text{argmax}}\; f\left(\mathbf{e}(\overrightarrow{v_{r}},\overrightarrow{v_{t}},\zeta),\mathbf{r}^{(i)}\right)$
\ENSURE Characteristic vector $\mathbf{e}(\overrightarrow{v_{r,i}},\overrightarrow{v_{t,i}},\zeta_i)$
\end{algorithmic}
\label{algo_joint}
\end{algorithm}

\noindent {\bf Sequential estimation.} Let us propose a new way of estimating the characteristic vector, taking profit of the analysis of the previous subsection. The idea is to exploit the parameters orthogonality, adopting a sequential strategy presented in algorithm~\ref{algo_seq}. In this algorithm, the first step (line 1) corresponds to the maximum likelihood estimation of the delay considering DoD and DoA as unknown nuisance parameters ($\mathbf{b}_k$ being the $k$-th vector of the standard basis of $\mathbb{R}^{N_t}$ and $\mathbf{c}_l$ being the $l$-th vector of the standard basis of $\mathbb{R}^{N_r}$), the second step (line 2) corresponds to the maximum likelihood estimation of the DoD considering a known delay and the DoA as an unknown nuisance parameter and finally the third step (line 3) corresponds to the maximum likelihood estimation of the DoA considering both delay and DoD are known. This strategy, if testing $N_\zeta$ delays, $N_{\overrightarrow{v_r}}$ DoAs and $N_{\overrightarrow{v_t}}$ DoDs requires $N_\zeta N_r N_t + N_{\overrightarrow{v_t}}N_r  + N_{\overrightarrow{v_r}} $ computations of the cost function $f(\mathbf{x},\mathbf{r}^{(i)})$. Thanks to the parameters orthogonality \cite{Cox1987}, this strategy is asymptotically (at high SNR or with a great number of measurements $N_m$) equivalent to the joint estimation, although it may be much less complex (provided $N_{\overrightarrow{v_t}} \gg N_t$ and $N_{\overrightarrow{v_r}} \gg N_r$). Note that this is true for any estimation order (the DoD or DoA can be estimated first also, without affecting the conclusions), so the order should be chosen so as to yield the lowest complexity. This sequential strategy is empirically assessed and compared to the classical joint strategy in section~\ref{sec:experiments}.
\begin{algorithm}[htb]
\caption{Sequential characteristic vector estimation}
\begin{algorithmic}[1] 
\REQUIRE Current residual $\mathbf{r}^{(i)}$, observation matrix $\mathbf{M}$
\STATE Estimate delay $\zeta_i$ based on $\mathbf{r}^{(i)}$ \\
$\zeta_i\leftarrow \underset{\zeta}{\text{argmax}}\; \sum_{k=1}^{N_t} \sum_{l=1}^{N_r}f\left(\mathbf{e}_f(\zeta)\otimes \mathbf{b}_k \otimes \mathbf{c}_l,\mathbf{r}^{(i)}\right)$
\STATE Estimate DoD $\overrightarrow{v_{t,i}}$ based on $\mathbf{r}^{(i)}$ and $\zeta_i$ \\
$\overrightarrow{v_{t,i}}\leftarrow \underset{\overrightarrow{v_{t}}}{\text{argmax}}\;  \sum_{l=1}^{N_r}f\left(\mathbf{e}_f(\zeta_i)\otimes \mathbf{e}_t(\overrightarrow{v_{t}})^*\otimes \mathbf{c}_l,\mathbf{r}^{(i)}\right)$
\STATE Estimate DoA $\overrightarrow{v_{r,i}}$ based on $\mathbf{r}^{(i)}$ , $\zeta_i$ and  $\overrightarrow{v_{t,i}}$\\
$\overrightarrow{v_{r,i}}\leftarrow \underset{\overrightarrow{v_{r}}}{\text{argmax}}\;  f\left(\mathbf{e}(\overrightarrow{v_{r}},\overrightarrow{v_{t,i}},\zeta_i),\mathbf{r}^{(i)}\right)$
\ENSURE Characteristic vector $\mathbf{e}(\overrightarrow{v_{r,i}},\overrightarrow{v_{t,i}},\zeta_i)$
\end{algorithmic}
\label{algo_seq}
\end{algorithm}

\section{Bias analysis}
\label{sec:bound_bias}

The main result of the previous section (theorem~\ref{thm:variance}) indicates that the variance of the studied class of estimators is at best proportional to the number of virtual paths $p$. This result may lead to choose $p$ as small as possible so as to minimize the variance. However, if $p$ is taken too small, $\mathbf{h}_{\mathcal{M}_p}$ may become an oversimplified version of the channel $\mathbf{h}$ leading to a high MSE due to a high model bias $\Vert\mathbf{h}-\mathbf{h}_{\mathcal{M}_p}\Vert_2^2$. There is a bias-variance tradeoff. How to set $p$ appropriately? Knowing that the number of physical paths $P$ is in general very large (up to several hundreds), is it possible to approximate it with a few virtual paths? This section studies these questions, trying to understand the mechanisms allowing to merge a large number of physical paths into much fewer virtual paths without incurring a large bias. 

Computing the bias defined in \eqref{eq:bias_variance} amounts to compute the projection $\mathbf{h}_{\mathcal{M}_p}$. Unfortunately, even considering a discretized set of candidates DoAs, DoDs and delays, this problem (which then becomes a sparse approximation problem) is NP-hard \cite{Tropp2010}. The projection can be approximated numerically using sparse recovery methods, as will be done in section~\ref{sec:experiments}. However the objective of this section is to study theoretically the bias and give an interpretable upper bound.   

\noindent {\bf Bound on the bias.} To do so, let us consider a simple situation in which $L$ physical paths are approximated by a single virtual path, in order to understand under which circumstances they can be merged. In that case, the physical channel is expressed
$$
\mathbf{h} = \sqrt{N_rN_tN_f}\sum\nolimits_{l=1}^L\beta_l\mathbf{e}(\overrightarrow{u_{r,l}},\overrightarrow{u_{t,l}},\tau_l),
$$
and the bias is bounded as 
$$\Vert \mathbf{h} - \mathbf{h}_{\mathcal{M}_1} \Vert_2 \leq \Vert \mathbf{h} - \tilde{\mathbf{h}} \Vert_2 $$
for any
$$
\tilde{\mathbf{h}} = \sqrt{N_rN_tN_f}\gamma\mathbf{e}(\overrightarrow{w_{r}},\overrightarrow{w_{t}},\xi),
$$
where $\overrightarrow{w_{r}}$,$\overrightarrow{w_{t}}$ and $\xi$ are the DoA, DoD and delay of the approximating virtual path.
This is true in particular considering the optimal coefficient
$$
\gamma_{\text{opt}} = \sum\nolimits_{l=1}^L\beta_l\mathbf{e}(\overrightarrow{w_{r}},\overrightarrow{w_{t}},\xi)^H\mathbf{e}(\overrightarrow{u_{r,l}},\overrightarrow{u_{t,l}},\tau_l)
$$
for which $\tilde{\mathbf{h}}$ is the orthogonal projection of the channel $\mathbf{h}$ onto the characteristic vector $\mathbf{e}(\overrightarrow{w_{r}},\overrightarrow{w_{t}},\xi)$.
With this optimal coefficient, introducing $\mathbf{e}\triangleq \mathbf{e}(\overrightarrow{w_{r}},\overrightarrow{w_{t}},\xi) $ and $\mathbf{e}_l\triangleq \mathbf{e}(\overrightarrow{u_{r,l}},\overrightarrow{u_{t,l}},\tau_l) $ in order to lighten notations, it follows
\begin{equation}
\def\arraystretch{1.8}
\begin{array}{rll}
\big\Vert\mathbf{h}-\tilde{\mathbf{h}}\big\Vert_2&= 
\sqrt{N_rN_tN_f}\Big\Vert\sum\nolimits_{l=1}^L \beta_l\left(\mathbf{e}_l - \mathbf{e}^H\mathbf{e}_l\mathbf{e}\right)\Big\Vert_2\\
&\leq 
\sqrt{N_rN_tN_f}\sum\nolimits_{l=1}^L \Big\Vert\beta_l\left(\mathbf{e}_l - \mathbf{e}^H\mathbf{e}_l\mathbf{e})\right)\Big\Vert_2\\
&= 
\sqrt{N_rN_tN_f}\sum\nolimits_{l=1}^L |\beta_l|\sqrt{1 - |\mathbf{e}^H\mathbf{e}_l|^2}.
\end{array}
\label{eq:ineqs_bias}
\end{equation}
This inequality highlights the fact that collinear characteristic vectors lead to small error. Said otherwise, physical paths with characteristic vectors collinear to the one of the approximating virtual path can be merged without accuracy loss.

Moreover, the inner product between characteristic vectors is expressed as the product of inner products between steering and delay vectors,
$$
\mathbf{e}^H\mathbf{e}_l = [\mathbf{e}_f(\xi)^H\mathbf{e}_f(\tau_l)][\mathbf{e}_t(\overrightarrow{u_{t,l}})^H\mathbf{e}_t(\overrightarrow{w_{t}})][\mathbf{e}_r(\overrightarrow{w_{r}})^H\mathbf{e}_t(\overrightarrow{u_{r,l}})].
$$
Analyzing jointly the three inner products yields the following approximation bound which is the main result of this section. 
\begin{theorem}
\label{thm:bias}
Provided $
\Vert \overrightarrow{u_{r,l}} -  \overrightarrow{w_{r}} \Vert_2< \frac{1}{\sqrt{2}\pi \frac{R_r}{\lambda}  }
$, 
$
\Vert \overrightarrow{u_{t,l}} -  \overrightarrow{w_{t}} \Vert_2< \frac{1}{\sqrt{2}\pi \frac{R_t}{\lambda}  }
$
and 
$
\vert \tau_l -  \xi \vert< \frac{1}{\frac{\pi}{\sqrt{2}} B  }
$
the bias is bounded by
$$
\begin{array}{l}
 \big\Vert\mathbf{h}-\mathbf{h}_{\mathcal{M}_1}\big\Vert_2 \leq \\
\sqrt{N_rN_tN_f}\sum\limits_{l=1}^L |\beta_l| \sqrt{1-(1-x_l)^2(1-y_l)^2(1-z_l)^2},
\end{array}
$$
where
\begin{itemize}[leftmargin=*] 
\item$
x_l = 2\pi^2(\tau_l-\xi)^2\tfrac{1}{N_f}\sum\nolimits_{i=1}^{N_f} f_i^2,
$
\item$
y_l = 2\pi^2\Vert \overrightarrow{u_{t,l}} -  \overrightarrow{w_{t}} \Vert_2^2\tfrac{1}{N_t}\sum\nolimits_{j=1}^{N_t} \frac{\Vert\overrightarrow{a_{t,j}}\Vert_2^2}{\lambda^2}\cos^2(\overrightarrow{a_{t,j}},\overrightarrow{u_{t,l}} -  \overrightarrow{w_{t}}),
$
\item$
z_l = 2\pi^2\Vert \overrightarrow{u_{r,l}} -  \overrightarrow{w_{r}} \Vert_2^2\tfrac{1}{N_r}\sum\nolimits_{k=1}^{N_r} \frac{\Vert\overrightarrow{a_{r,k}}\Vert_2^2}{\lambda^2}\cos^2(\overrightarrow{a_{r,k}},\overrightarrow{u_{r,k}} -  \overrightarrow{w_{r}}).
$
\end{itemize}
\end{theorem}
The theorem is proven in appendix~\ref{app:proof_thm_bias}, in order to keep the flow of the paper.

\noindent {\bf Interpretations.} 
This bound yields quite intuitive results that can be nicely interpreted:
\begin{itemize}[leftmargin=*]
\item First of all, physical paths that are close to the approximating virtual path in the delay, DoD and DoA domains can be merged. Indeed, if the quantities $(\tau_l-\xi)^2$, $\Vert \overrightarrow{u_{t,l}} -  \overrightarrow{w_{t}} \Vert_2^2$ and $\Vert \overrightarrow{u_{r,l}} -  \overrightarrow{w_{r}} \Vert_2^2$ are small the $l$-th term of the bound is small and the $l$-th path can be merged with low approximation error. It is important to notice that the physical path and the virtual path have to be close in the three domains, because one domain in which they are far apart is sufficient to get almost orthogonal characteristic vectors.
\item Second, the discrimination power of the system is taken into account by the bound. A larger bandwidth or larger arrays lead to more difficulty to merge physical paths. Indeed, if the quantities $\sum\nolimits_{i=1}^{N_f} f_i^2$, $\sum\nolimits_{j=1}^{N_t} \frac{\Vert\overrightarrow{a_{t,j}}\Vert_2^2}{\lambda^2}$ and $\sum\nolimits_{k=1}^{N_r} \frac{\Vert\overrightarrow{a_{r,k}}\Vert_2^2}{\lambda^2}$ increase the approximation error also increases. On the other hand, if there is for example only one antenna at the receiver, a physical path that has a close delay and DoD with the approximating virtual path can be merged irrespective of its DoA, because the system has then no discrimination power in the DoA domain.
\item Finally, the ability to merge paths depends also on the orientation of the transmit and receive antenna arrays. Some paths can be merged even if they are far from the virtual path in the DoD and DoA domains, if their direction differences lie in a direction in which the array is insensitive. This can be seen by the presence of the quantities $\cos^2(\overrightarrow{a_{t,j}},\overrightarrow{u_{t,l}} -  \overrightarrow{w_{t}})$ and $\cos^2(\overrightarrow{a_{r,k}},\overrightarrow{u_{r,k}} -  \overrightarrow{w_{r}})$ in the bound. for example for a ULA aligned with the $z$-axis, a path that has the same elevation as the approximating virtual path will be perfectly mergeable irrespective of  its azimuth, because the ULA has sensitivity only in one direction (the cosine will be null in this case). 
\end{itemize}

 The proposed bound lends itself to nice interpretations as shown above, but may be loose because of the use of the triangle inequality. A way of possibly improving the bound is evoked in appendix~\ref{app:improving_the_bound}.

\begin{figure*}[h]
\begin{subfigure}[b]{0.333\textwidth}
\includegraphics[width=\columnwidth]{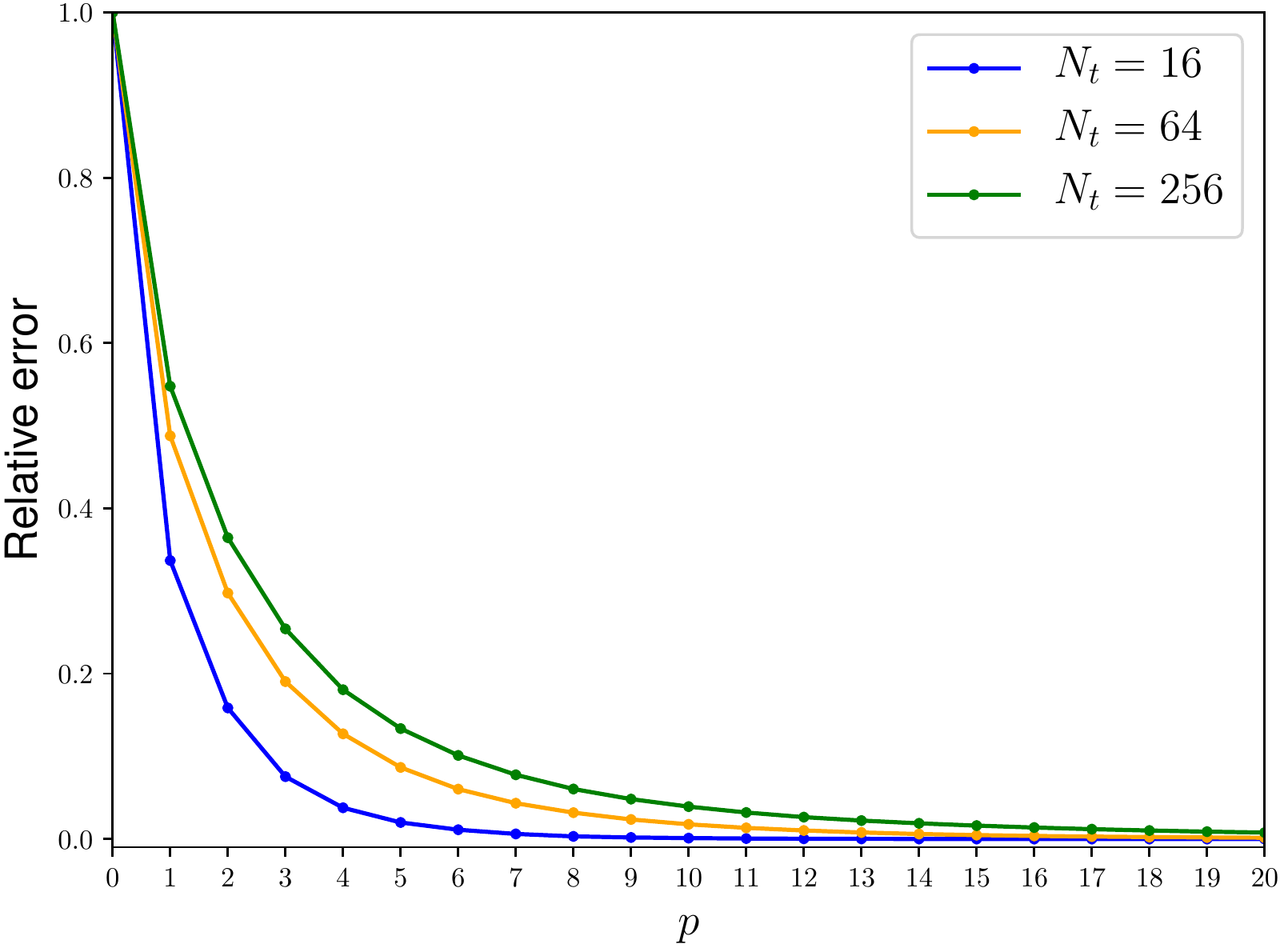}
\caption{}
\label{fig:antennas}
\end{subfigure}
\begin{subfigure}[b]{0.333\textwidth}
\includegraphics[width=\columnwidth]{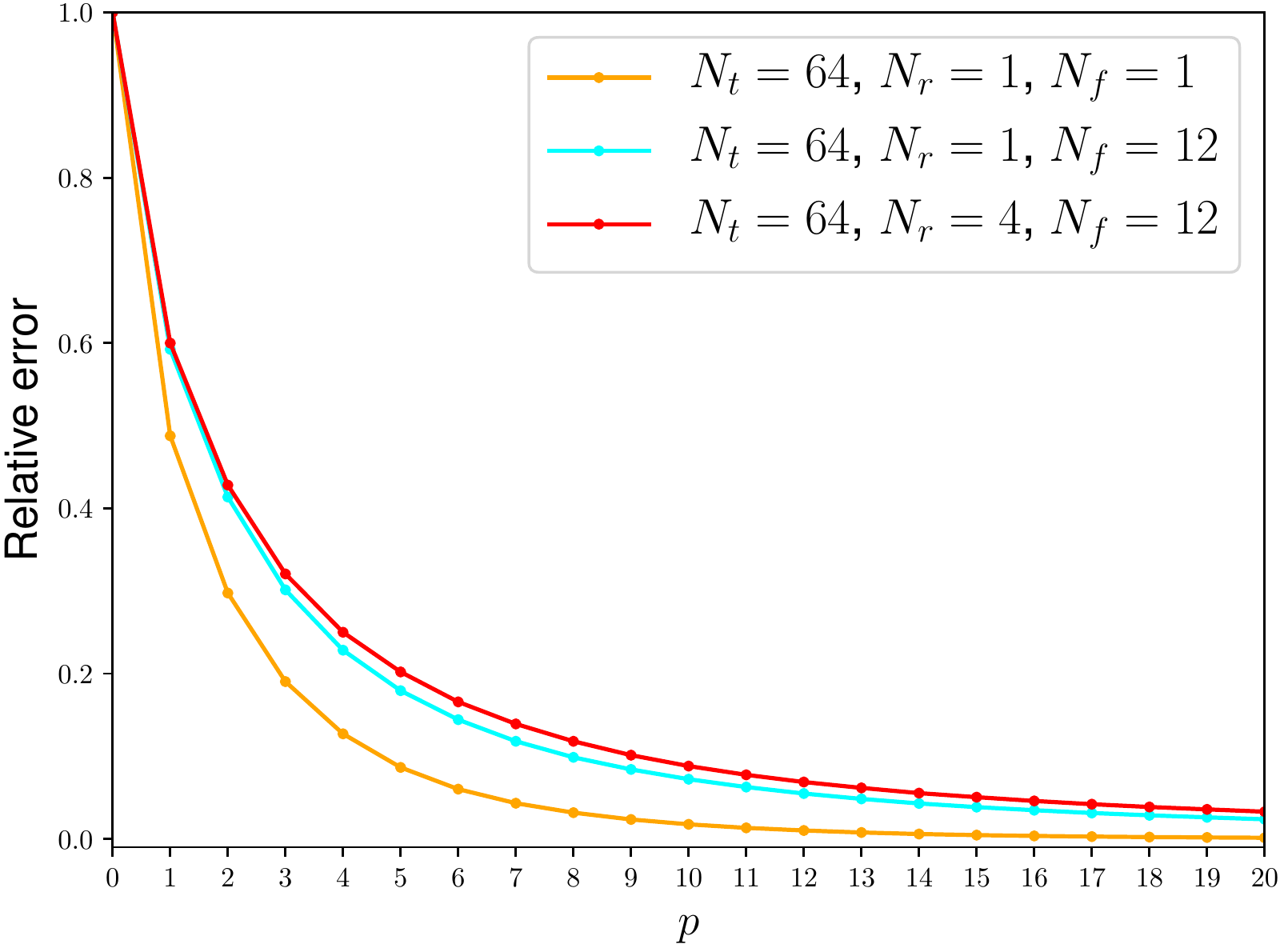}
\caption{}
\label{fig:params}
\end{subfigure}
\begin{subfigure}[b]{0.333\textwidth}
\includegraphics[width=\columnwidth]{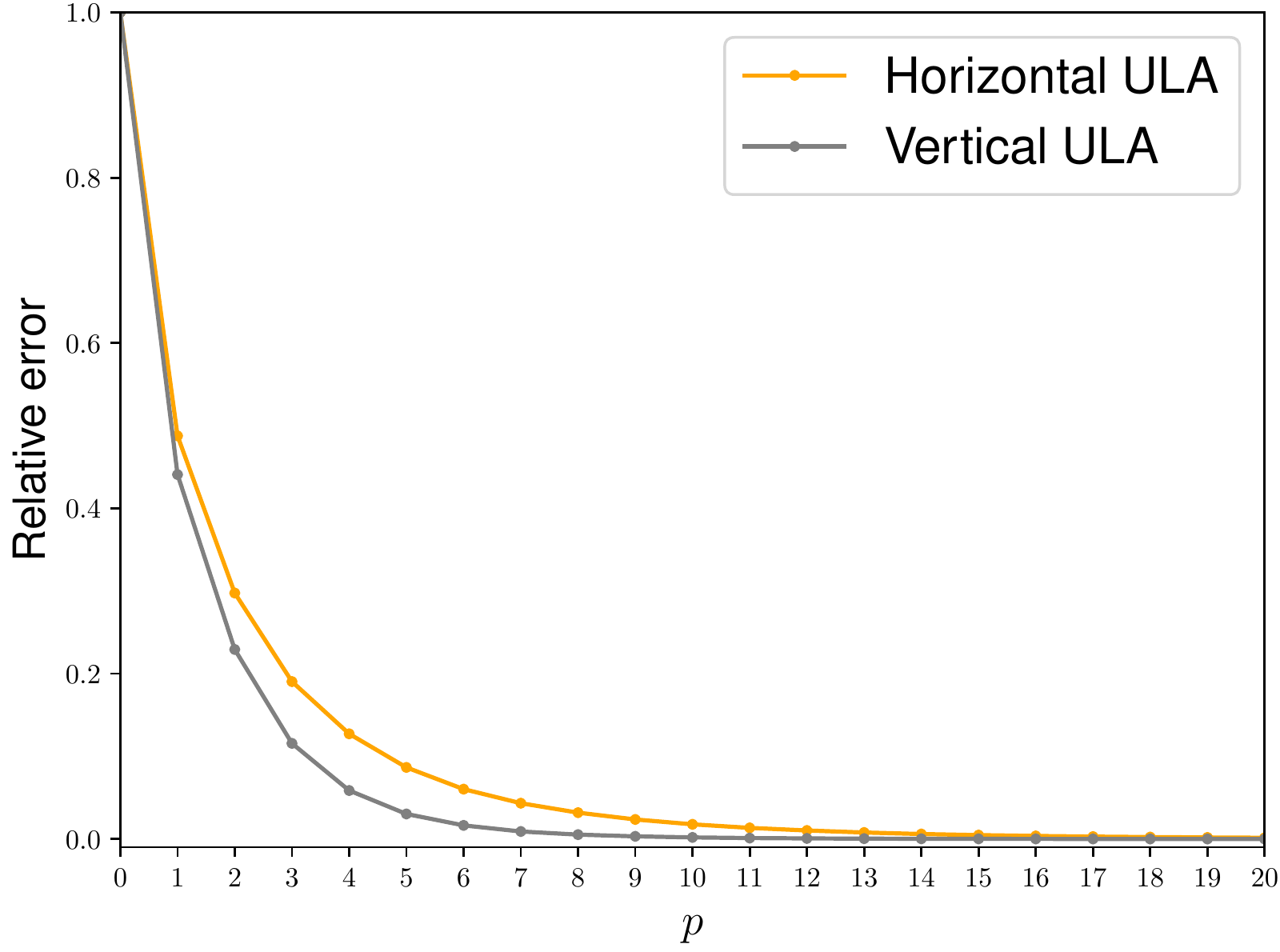}
\caption{}
\label{fig:orientation}
\end{subfigure}
\caption{Bias assessment in several configurations with a varying number of virtual paths $p$.}
\label{fig:bias}
\end{figure*}
\section{Experiments}
\label{sec:experiments}
The objective of this section is to assess empirically the mathematical developments of the previous sections. The model bias is first computed with a varying number of virtual paths. Then, the joint and sequential estimation strategies (algorithms~\ref{algo_joint} and \ref{algo_seq}) are compared, showing a bias-variance tradeoff. All experiments are done using realistic channels generated with help of the NYUSIM channel simulator \cite{Sun2017}, in a millimeter wave massive MIMO context. In particular, the central frequency is set to $f_c=28$\,GHz and the distance between transmitter and receiver to $d=30$\,m to obtain the DoDs, DoAs, delays, gains and phases of each path. The channel matrix is then obtained from \eqref{eq:physical_channel} (with the total number of physical paths $P$ being random and ranging from fifty to a hundred). All results shown in this section are averages over one hundred channel realizations.

\subsection{Empirical evaluation of the model bias}
\label{ssec:expe_bias}

Computing the model bias $\Vert\mathbf{h}-\mathbf{h}_{\mathcal{M}_p}\Vert_2^2$ requires to compute the projection onto the model $\mathbf{h}_{\mathcal{M}_p}=\text{proj}_{\mathcal{M}_p}(\mathbf{h})$. This is unfortunately a NP-hard problem even with discretized directions and delays \cite{Tropp2010}. It is nevertheless possible to compute an approximation $\hat{\mathbf{h}}_{\mathcal{M}_p}$ by using algorithm~\ref{algo_greedy} called directly on $\mathbf{h}$ (in perfect observation conditions and without noise). The relative bias is then approximated by the relative error 
$$
\frac{\Vert\mathbf{h}-\hat{\mathbf{h}}_{\mathcal{M}_p}\Vert_2^2}{\Vert\mathbf{h}\Vert_2^2}.
$$

For this first set of experiments, let us compute this relative error in various configurations (varying $p$, $N_t$, $N_r$, $N_f$ and the orientation of the antenna array).
 To do so, algorithm~\ref{algo_greedy} is used in conjunction with the joint characteristic vector estimation (algorithm~\ref{algo_joint}), for which the optimization problem is solved by exhaustive testing of $N_\zeta$ delays, $N_{\overrightarrow{v_r}}$ DoAs and $N_{\overrightarrow{v_t}}$ DoDs (evenly sampling the three domains).
In order to simplify the analysis, let us have a single oversampling parameter $S$ controlling the number of tests. This amounts to take $N_\zeta = SN_f$ if $N_f >1$, $N_\zeta = 1$ otherwise (the delay has no influence with $N_f=1$), and setting similarly $N_{\overrightarrow{v_r}}$ and $N_{\overrightarrow{v_t}}$. The value $S=6$ is taken here (it was found empirically that testing more values does not improve the result, the impact or reducing $S$ is assessed in the next subsection). Results on average over $100$ NYUSIM channel realizations are shown on figure~\ref{fig:bias}. Several comments are in order:
\begin{itemize}[leftmargin=*]
\item On all three subfigures, the error decreases quickly when the number of virtual paths $p$ increases. This shows that even a large number of physical paths can indeed be merged into much fewer virtual paths with little accuracy loss. This justifies the use of channel estimators based on physical models taking the form of \eqref{eq:estimator_channel} for massive MIMO in the millimeter wave band.
\item On figure~\ref{fig:antennas}, the influence of the number of antennas on the bias is assessed. To do so, let us consider an uniform linear array (ULA) aligned with the $x$-axis with half-wavelength separated antennas with a varying number of transmit antennas $N_t \in \{16,64,256\}$ at the transmitter, a single antenna receiver ($N_r = 1$) and a single subcarrier ($N_f=1$). First, notice that taking more virtual paths (a greater $p$) leads to lower error, as expected since each physical path is then closer in average to a virtual path. This could be predicted from the bound of theorem~\ref{thm:bias}, in which adding virtual paths amounts to reduce the quantity $\Vert \overrightarrow{u_{t,l}} -  \overrightarrow{w_{t}} \Vert_2^2$. This conclusion holds also for figures~\ref{fig:params} and \ref{fig:orientation}. Second, more antennas require more virtual paths to attain the same error, because the discrimination power of the antenna array is higher, as was predicted by the quantity $\sum\nolimits_{j=1}^{N_t} \frac{\Vert\overrightarrow{a_{t,j}}\Vert_2^2}{\lambda^2}$ in the bound of theorem~\ref{thm:bias}.

\item Let us now consider a fixed number of transmit antennas ($64$) and add several receive antenna and several subcarriers. Results of this experiment are shown on figure~\ref{fig:params}. It is clear that considering $12$ subcarriers (with a $15$ MHz spacing between adjacent subcarriers) leads to a higher discrimination power, which results in higher error for a fixed number of virtual paths. This is also the case when a ULA with $4$ receive antennas is considered, but the difference is subtle, since the DoA discrimination power of an array with $4$ antennas is quite low. These observations are nicely interpreted with help of the bound of theorem~\ref{thm:bias}, because the quantities $\sum\nolimits_{i=1}^{N_f} f_i^2$ and $\sum\nolimits_{k=1}^{N_r} \frac{\Vert\overrightarrow{a_{r,k}}\Vert_2^2}{\lambda^2}$ are not null with several subcarriers and receive antennas, whereas they were for the previous experiment.

\item Finally, changing the orientation of the antenna array also has an influence on the bias, as evidenced on figure~\ref{fig:orientation}. Indeed, taking a ULA aligned with the $z$-axis (a vertical ULA) allows to  get a lower error for a fixed number of virtual paths compared to the horizontal ULA aligned with the $x$-axis. This is because the physical paths generated by NYUSIM have mostly similar elevations, they differ mainly for their azimuth. This observation corresponds in the bound of theorem~\ref{thm:bias} to $\cos^2(\overrightarrow{a_{t,j}},\overrightarrow{u_{t,l}} -  \overrightarrow{w_{t}})$ that is small in most cases for the vertical ULA, since it has discrimination power only on the elevation, and not on the azimuth (this is the opposite for the horizontal ULA).
\end{itemize}

\begin{figure*}[htb]
\begin{subfigure}[b]{0.333\textwidth}
\includegraphics[width=\columnwidth]{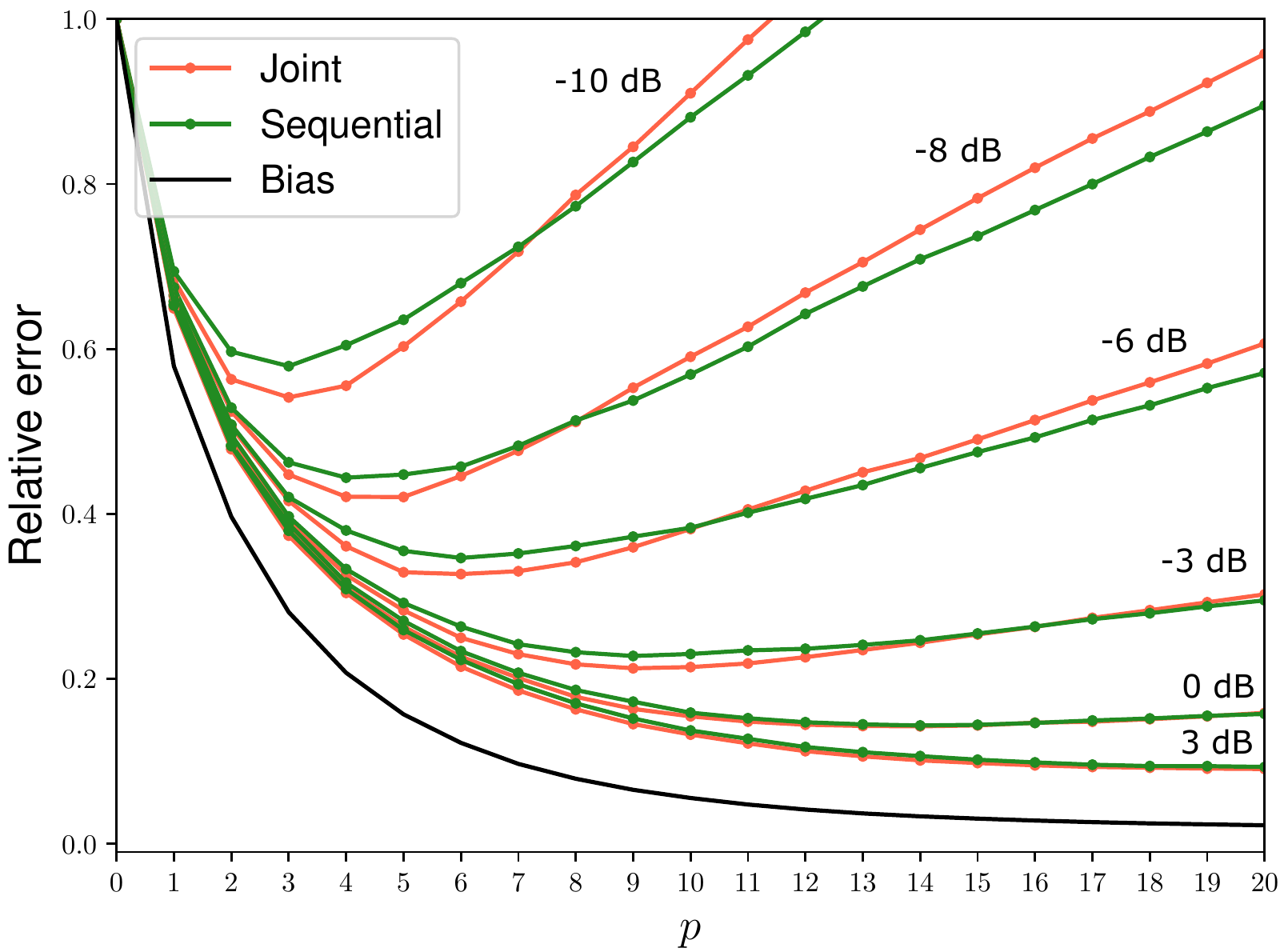}
\caption{$S = 2$, $\frac{T_\text{joint}}{T_\text{seq}} = 1.70$}
\label{fig:s2}
\end{subfigure}
\begin{subfigure}[b]{0.333\textwidth}
\includegraphics[width=\columnwidth]{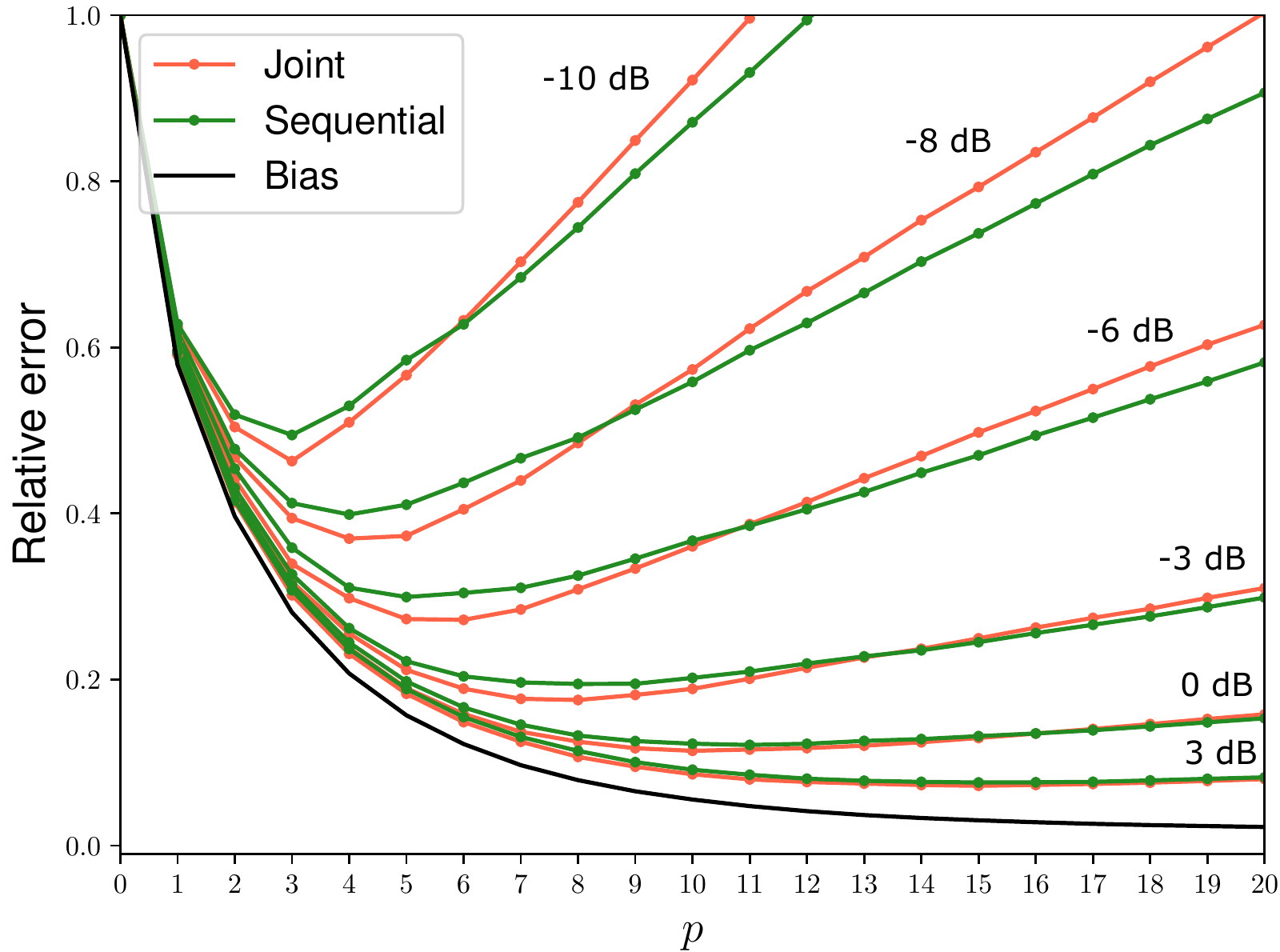}
\caption{$S = 4$, $\frac{T_\text{joint}}{T_\text{seq}} = 3.56$}
\label{fig:s4}
\end{subfigure}
\begin{subfigure}[b]{0.333\textwidth}
\includegraphics[width=\columnwidth]{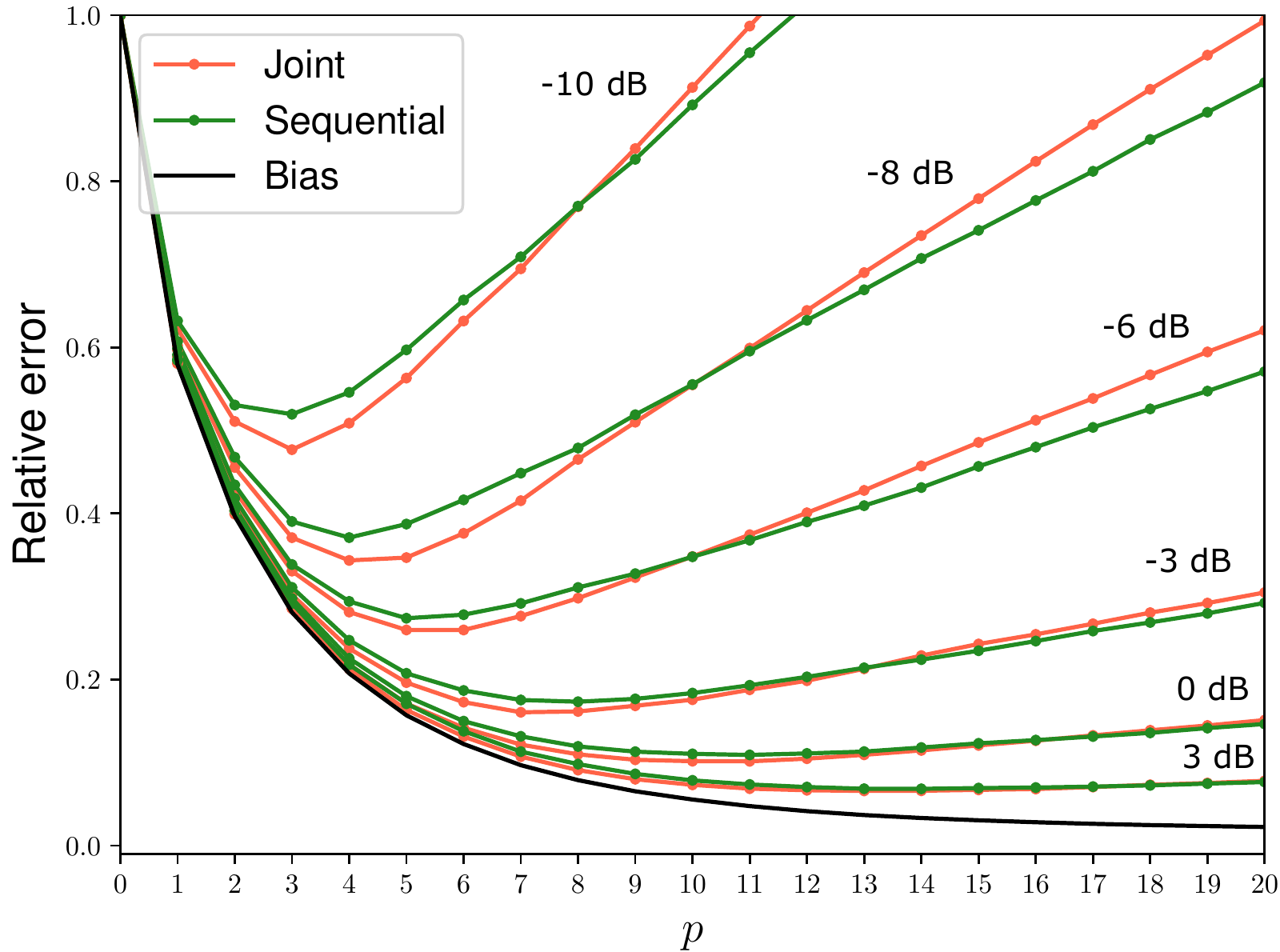}
\caption{$S = 6$, $\frac{T_\text{joint}}{T_\text{seq}} = 5.55$}
\label{fig:s6}
\end{subfigure}
\caption{Comparison of the joint and sequential strategies for several oversampling factors $S$ and several SNRs.}
\label{fig:compmethods}
\end{figure*}

\subsection{Comparison of joint and sequential strategies}
\label{ssec:expe_compmethods}

The objective of this subsection is to compare the joint and sequential estimation strategies (algorithms~\ref{algo_joint} and \ref{algo_seq}) for several signal to noise ratios (SNR). To do so, let us consider a system with $N_t = 64$ transmit antennas (arranged as an ULA with half-wavelength separated antennas), a single receive antenna ($N_r=1$) and $N_f=12$ subcarriers (with a $15$ MHz spacing between adjacent subcarriers). When the sequential characteristic vector estimation (algorithm~\ref{algo_seq}) is used, the DoD is estimated first and then the delay (the DoA being irrelevant since $N_r=1$). The observation matrix $\mathbf{M}$ is taken as the identity (the objective here being to study the channel model and not pilot sequences). The greedy channel estimation algorithm is used on noisy observations taking the form of \eqref{eq:observation} (with various noise levels) to get a channel estimate $\hat{\mathbf{h}}$. The oversampling factor $S$ defined in the previous subsection is taken in $\{2,4,6\}$. The performance is measured by the relative error 
$$
\frac{\Vert\mathbf{h}-\hat{\mathbf{h}}\Vert_2^2}{\Vert\mathbf{h}\Vert_2^2}.
$$

Results are shown on figure~\ref{fig:compmethods}, where each subfigure corresponds to a different oversampling factor $S$. Note that the approximate bias (computed with $S=6$) is shown on the three plots to serve as a reference. Several comments are in order:
\begin{itemize}[leftmargin=*]
\item First, notice that for both characteristic vector estimation methods, the error is close to the bias for a small $p$, and then linearly increasing for a large $p$ (error dominated by the variance). This is a bias-variance tradeoff. It is interesting to notice that the optimal number of virtual paths increases with the SNR (for both methods) and stays relatively small (no more than a dozen, which yields around $50$ real parameters to estimate), whereas the channel vector has $768$ complex entries (which yields $1536$ real parameters to estimate if not using a parametric model). For a large $p$, the slope at which the error increases is inversely proportional to the SNR, which was predicted by corollary~\ref{cor:variance}.

\item Second, the joint strategy (algorithm~\ref{algo_joint}) performs better than the sequential strategy (algorithm~\ref{algo_seq}) for all SNRs (at the optimal $p$). However, the difference becomes very small at high SNR. For example, the difference is approximately $3\%$ for a SNR of $-8$ dB and  only approximately $0.5\%$ for a SNR of $0$ dB. This is in total agreement with the parameter orthogonality between DoD and delay evidenced in section~\ref{sec:variance}, which implies asymptotic independence.

\item Finally, one can notice that increasing the oversampling factor $S$ improves accuracy, but also increases complexity. Indeed, the joint strategy complexity is $S^2N_tN_f$ whereas the sequential strategy complexity is $SN_tN_f + SN_f$ (it is roughly $S$ times less complex). This implies that the complexity advantage of the sequential strategy increases linearly with $S$ (it would increase quadratically with several receive antennas and the estimation of the DoA). This is verified empirically by the ratio of the average joint estimation time $T_{\text{joint}}$ and the average sequential estimation time $T_{\text{seq}}$ shown below each figure, which indeed increases with $S$.
\end{itemize}

In summary, the sequential strategy was empirically found to be almost as good as the joint strategy, whereas it is much less complex. Moreover, its complexity advantage is higher if a high accuracy is sought for. This shows that such sequential characteristic vector estimation strategies are promising for greedy channel estimation methods.

\section{Conclusion}
In this paper, the performance of MIMO wideband channel estimators using a physical model was theoretically studied. To do so, the mean squared error (MSE) of channel estimation was decomposed in to a bias and a variance term. The variance was studied with help of the Cram\'er-Rao bound, shown to be proportional to the number of considered virtual paths, provided a condition on the observation matrix $C_{\text{opt}}$ is fulfilled. Moreover, computing the Fisher information matrix (FIM) allowed to discover that DoD, DoA and delay are orthogonal parameters, leading to the design of a computationally efficient  sequential estimation algorithm (algorithm~\ref{algo_seq}) with asymptotic optimality properties. Then, the bias term was bounded by a quantity lending itself to nice interpretations, depending both on the propagation properties of the channel, on the antenna arrays geometries and on the position of the subcarriers.

The aforementioned mathematical developments were then assessed experimentally. First, the bias was approximated in various configurations, verifying the predictions of the computed bound, and furthermore showing that few virtual paths are sufficient to approximate well realistic channels in the millimeter wave band. The sequential estimation algorithm was then compared to the classical joint one. It was found that from moderately high SNR it performed as well as the joint one, whereas being much more computationally efficient. This shows it is a promising alternative and should be considered for system design.

In the future, it would be of great interest to study practical designs of observation matrices that take into account practical constraints (for example hybrid systems) while fulfilling exactly or approximately the optimal observation condition $C_{\text{opt}}$ and requiring a reasonable amount of resources (with a number of linear measurements $N_m$ much smaller than $N_tN_rN_f$). On a more technical side, improving the bound on the bias proposed here may be possible, as mentioned in appendix~\ref{app:improving_the_bound}.

\appendix

\subsection{Proof of theorem~\ref{thm:variance}}
\label{app:proof_thm_variance}

Starting from \eqref{eq:FIM_gen} and denoting $\mathbf{A} \triangleq \mathbf{M}^H\boldsymbol{\Sigma}^{-1}\mathbf{M}$, the CRB can be re-expressed using only real matrices by introducing
$$
\bar{\mathbf{D}} \triangleq \left( \begin{array}{c}
\mathfrak{Re}\{\frac{\partial \mathbf{h}_{\mathcal{M}_p}}{\partial \boldsymbol{\theta}}\} \\
\mathfrak{Im}\{\frac{\partial \mathbf{h}_{\mathcal{M}_p}}{\partial \boldsymbol{\theta}}\}
\end{array} \right),\quad 
\bar{\mathbf{A}} \triangleq \left(\begin{array}{cc}
\mathfrak{Re}\{\mathbf{A}\} & -\mathfrak{Im}\{\mathbf{A}\} \\
\mathfrak{Im}\{\mathbf{A}\} & \mathfrak{Re}\{\mathbf{A}\}
\end{array}\right), 
$$
so that
$$
\mathbf{I}(\boldsymbol{\theta}) = 2\bar{\mathbf{D}}^T\bar{\mathbf{A}}\bar{\mathbf{D}}
$$
is verified immediately. This yields the following expression for the CRB
$$
\text{CRB} = \frac{1}{2}\text{Tr}\left[ \frac{\partial \mathbf{h}_{\mathcal{M}_p}}{\partial \boldsymbol{\theta}}(\bar{\mathbf{D}}^T\bar{\mathbf{A}}\bar{\mathbf{D}})^{-1} \frac{\partial \mathbf{h}_{\mathcal{M}_p}}{\partial \boldsymbol{\theta}}^H\right].
$$
Moreover, since for any symmetric real matrix $\mathbf{F}$ and any complex matrix $\mathbf{E}$,
$$
\text{Tr}\left[ \mathbf{EFE}^H \right] = \text{Tr}\left[ \begin{pmatrix}
\mathfrak{Re}\{\mathbf{E}\} \\
\mathfrak{Im}\{\mathbf{E}\}
\end{pmatrix}
\mathbf{F} 
\begin{pmatrix}
\mathfrak{Re}\{\mathbf{E}\}^T &
\mathfrak{Im}\{\mathbf{E}\}^T
\end{pmatrix}\right],
$$
it follows
\begin{equation}
\begin{array}{rl}
\text{CRB} &= \frac{1}{2}\text{Tr}\left[ \bar{\mathbf{D}}(\bar{\mathbf{D}}^T\bar{\mathbf{A}}\bar{\mathbf{D}})^{-1} \bar{\mathbf{D}}^T\right]\\
& = \frac{1}{2\Vert \bar{\mathbf{A}} \Vert_2}\text{Tr}\left[ \bar{\mathbf{D}}(\bar{\mathbf{D}}^T\frac{\bar{\mathbf{A}}}{\Vert \bar{\mathbf{A}} \Vert_2}\bar{\mathbf{D}})^{-1} \bar{\mathbf{D}}^T\right].
\end{array}
\end{equation}
The normalized matrix $\frac{\bar{\mathbf{A}}}{\Vert \bar{\mathbf{A}} \Vert_2}$ is symmetric positive semidefinite and has all its eigenvalues upper bounded by one so that
$$\frac{\bar{\mathbf{A}}}{\Vert \bar{\mathbf{A}} \Vert_2} \preceq \mathbf{Id},$$ which yields 
$$\bar{\mathbf{D}}^T\frac{\bar{\mathbf{A}}}{\Vert \bar{\mathbf{A}} \Vert_2}\bar{\mathbf{D}} \preceq \bar{\mathbf{D}}^T\bar{\mathbf{D}}.$$ 
Using \cite[Theorem 4.3]{Baksalary1989} and assuming the model is identifiable (so that the inversion is possible), it follows 
$$(\bar{\mathbf{D}}^T\frac{\bar{\mathbf{A}}}{\Vert \bar{\mathbf{A}} \Vert_2}\bar{\mathbf{D}})^{-1} \succeq (\bar{\mathbf{D}}^T\bar{\mathbf{D}})^{-1}, $$
and finally $$\bar{\mathbf{D}}(\bar{\mathbf{D}}^T\frac{\bar{\mathbf{A}}}{\Vert \bar{\mathbf{A}} \Vert_2}\bar{\mathbf{D}})^{-1}\bar{\mathbf{D}}^T \succeq \bar{\mathbf{D}}(\bar{\mathbf{D}}^T\bar{\mathbf{D}})^{-1}\bar{\mathbf{D}}^T, $$ 
which gives 
$$\text{Tr}\left[\bar{\mathbf{D}}(\bar{\mathbf{D}}^T\frac{\bar{\mathbf{A}}}{\Vert \bar{\mathbf{A}} \Vert_2}\bar{\mathbf{D}})^{-1}\bar{\mathbf{D}}^T\right] \geq \text{Tr}\left[ \bar{\mathbf{D}}(\bar{\mathbf{D}}^T\bar{\mathbf{D}})^{-1}\bar{\mathbf{D}}^T\right] = N_{\boldsymbol{\theta}}. $$
Dividing both sides by $2\Vert \bar{\mathbf{A}} \Vert_2$ and noticing that the matrices $\mathbf{A}$ and $\bar{\mathbf{A}}$ have the same eigenvalues (even though with a doubled multiplicity for $\bar{\mathbf{A}}$) so that $\left\Vert \mathbf{A} \right\Vert_2 = \left\Vert \bar{\mathbf{A}} \right\Vert_2$, \eqref{eq:thm_variance} is obtained.

\subsection{Proof of theorem~\ref{thm:bias}}
\label{app:proof_thm_bias}
The inner product of characteristic vectors involved in \eqref{eq:ineqs_bias} can be expressed as
$$
\mathbf{e}^H\mathbf{e}_l = [\mathbf{e}_f(\xi)^H\mathbf{e}_f(\tau_l)][\mathbf{e}_t(\overrightarrow{u_{t,l}})^H\mathbf{e}_t(\overrightarrow{w_{t}})][\mathbf{e}_r(\overrightarrow{w_{r}})^H\mathbf{e}_t(\overrightarrow{u_{r,l}})],
$$
which is a product of three inner products, that can all be expressed in the general form
$$
\frac{1}{N}\sum_{n=1}^N \mathrm{e}^{-\mathrm{j}C_n},
$$
where
\begin{itemize}
 \item taking $C_n = 2\pi f_n(\tau_l - \xi)$ and $N=N_f$ yields the delay vectors inner product $\mathbf{e}_f(\xi)^H\mathbf{e}_f(\tau_l)$,
\item taking $C_n = \frac{2\pi}{\lambda} \overrightarrow{a_{t,n}}.(\overrightarrow{w_t} - \overrightarrow{u_{t,l}} )$ and $N=N_t$ yields the transmit steering vectors inner product $\mathbf{e}_t(\overrightarrow{u_{t,l}})^H\mathbf{e}_t(\overrightarrow{w_{t}})$,
\item taking $C_n = \frac{2\pi}{\lambda} \overrightarrow{a_{r,n}}.(\overrightarrow{u_{r,l}} - \overrightarrow{w_r})$ and
 $N=N_r$ yields the receive steering vectors inner product $\mathbf{e}_r(\overrightarrow{w_{r}})^H\mathbf{e}_r(\overrightarrow{u_{r,l}})$.
\end{itemize}
This general expression allows to study these three inner products jointly.

Using the series representation of the exponential $
\mathrm{e}^x = \sum_{k=0}^{+\infty}\frac{x^k}{k!}
$, one gets
$$
\frac{1}{N}\sum_{n=1}^N \mathrm{e}^{-\mathrm{j}C_n} = \frac{1}{N}\sum_{n=1}^N\sum_{k=0}^{+\infty}\frac{(-j)^kC_n^k}{k!}.
$$
The antennas at both the transmitter and receiver being located with respect to the centroid of the array and the frequencies being expressed with respect to the central frequency, one gets $\sum_{n=1}^{N}C_n=0$, which implies that the term $k=1$ of the series is null in all three cases, leading to

$$
\frac{1}{N}\sum_{n=1}^N \mathrm{e}^{-\mathrm{j}C_n} = 1+\sum_{k=2}^{+\infty}\frac{\big(-\mathrm{j}\big)^k }{k!}\frac{1}{N}\sum_{n=1}^NC_n^k. 
$$
Noticing that $C_n^k\geq 0$ for $k$ even, one gets that the real part of the series is an alternating series. Its terms are decreasing in magnitude if
$$
C_n^k>\frac{k!}{(k+2)!}C_n^{(k+2)}, \quad \forall k,n.
$$
This is the case for any $k$ if it is fulfilled for $k=0$, leading to the sufficient condition 
$$
C_n^{2}<2, \quad \forall n.
$$
Specializing this condition to the three values $C_n$ can take allows to get the three sufficient conditions of the theorem: 
\begin{itemize}
\item $\vert \tau_l -  \xi \vert< \frac{1}{\frac{\pi}{\sqrt{2}} B  }$,
\item$\Vert \overrightarrow{u_{r,l}} -  \overrightarrow{w_{r}} \Vert_2< \frac{1}{\sqrt{2}\pi \frac{R_r}{\lambda}  }$,
\item $\Vert \overrightarrow{u_{t,l}} -  \overrightarrow{w_{t}} \Vert_2< \frac{1}{\sqrt{2}\pi \frac{R_t}{\lambda}  }$.
\end{itemize}
If they are fulfilled, the real part of the series can be bounded, 
$$
\mathfrak{Re}\left\{\frac{1}{N}\sum_{n=1}^N \mathrm{e}^{-\mathrm{j}C_n}\right\} \geq 1- \frac{1}{2N}\sum_{n=1}^N C_n^2,
$$
which implies 
$$
\left|\frac{1}{N}\sum_{n=1}^N \mathrm{e}^{-\mathrm{j}C_n}\right| \geq 1- \frac{1}{2N}\sum_{n=1}^N C_n^2.
$$
Replacing $C_n$ and $N$ by their respective values for the three inner products gives the theorem.

\subsection{Improving the bound on the bias}
\label{app:improving_the_bound}
The bound of this paper is obtained applying the triangle inequality, which may lead to a loose bound. Instead, one could write
$$
\begin{array}{rl}
\big\Vert\mathbf{h}-\tilde{\mathbf{h}}\big\Vert_2&= 
\sqrt{N_rN_tN_f}\Big\Vert\sum\nolimits_{l=1}^L \beta_l\left(\mathbf{e}_l - \mathbf{e}^H\mathbf{e}_l\mathbf{e}\right)\Big\Vert_2\\
& = \boldsymbol{\beta}^H\mathbf{Q}\boldsymbol{\beta}
\end{array}
$$
where $\boldsymbol{\beta} = (\beta_1,\dots,\beta_L)^T$ and $\mathbf{Q} \in \mathbb{C}^{L\times L}$
with
$$
q_{ij} = \mathbf{e}_i^H\mathbf{e}_j - \mathbf{e}_i^H\mathbf{e}\mathbf{e}^H\mathbf{e}_j.
$$
Studying the properties of the matrix $\mathbf{Q}$ may lead to a tighter bound, but is not guaranteed to lead to as easily interpretable results.

\bibliographystyle{IEEEtran}
\bibliography{biblio_mimo}

%
\IEEEpeerreviewmaketitle

\begin{IEEEbiography}[{\includegraphics[width=1in,height=1.25in,viewport=70 50 420 487.5,clip,keepaspectratio]{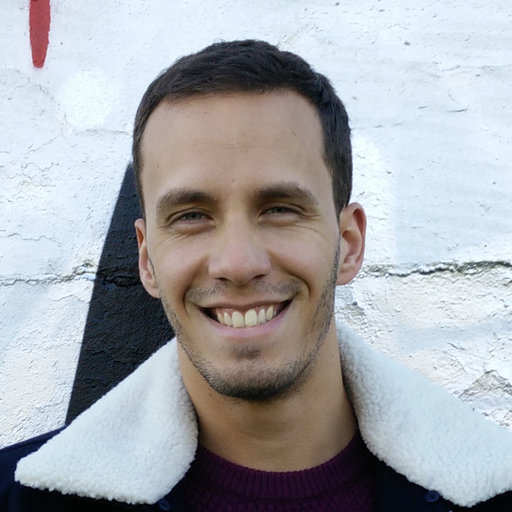}}]{Luc Le Magoarou} is a postdoctoral researcher at b\raisebox{0.2mm}{\scalebox{0.7}{\textbf{$<>$}}}com (Rennes, France). He received the Ph.D. in signal processing and the M.Sc. in electrical engineering, both from the National Institute of Applied Sciences (INSA Rennes, France), in 2016 and 2013 respectively. During his Ph.D., he was with Inria (Rennes, France), in the PANAMA research group. His main research interests are signal processing and machine learning, currently applied to MIMO communication systems. 
\end{IEEEbiography}
\begin{IEEEbiography}[{\includegraphics[width=1in,height=1.25in,viewport=50 50 540 662.5,clip,keepaspectratio]{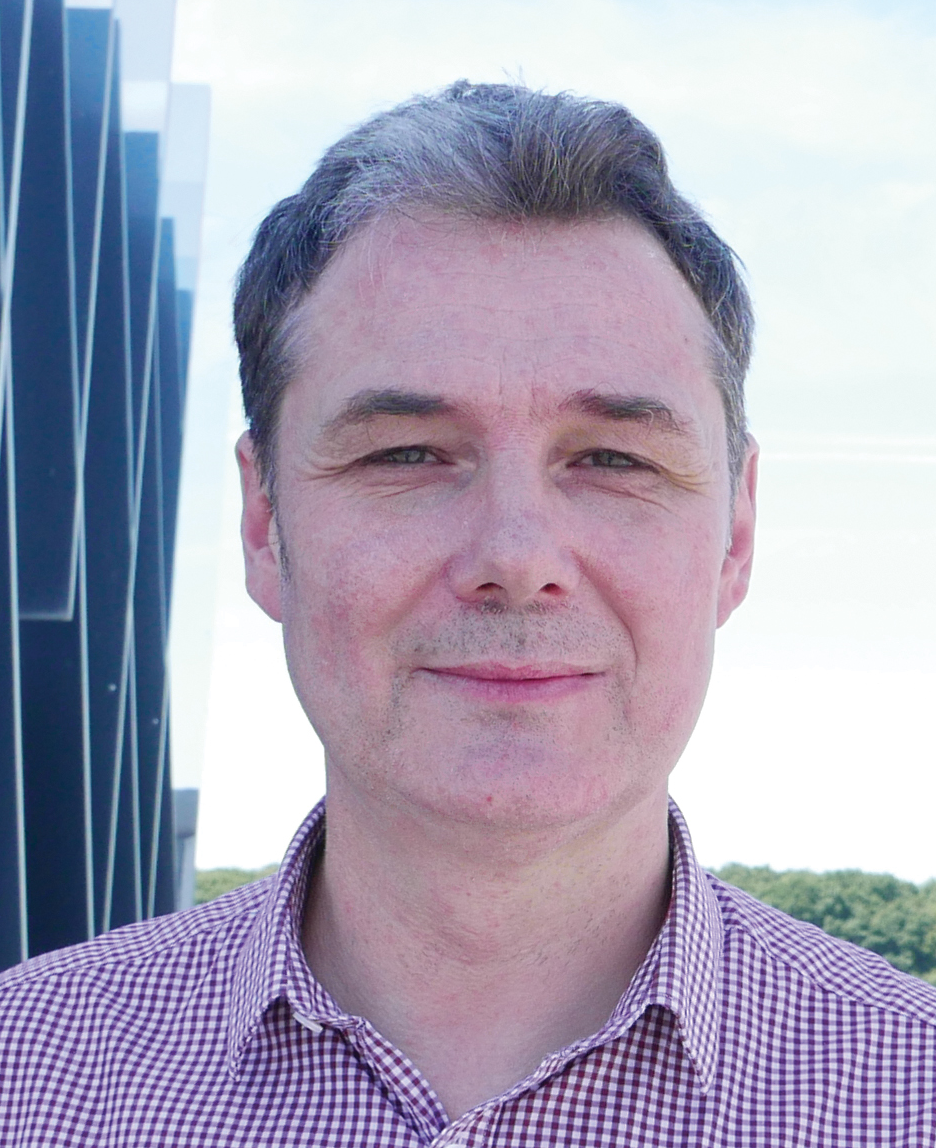}}]{St\'ephane Paquelet} received the B.Sc. degree
from the Ecole Polytechnique, Paris, France, in 1996, and the M.Sc. degree
from Telecom Paris, Paris, France, in 1998. He successively worked in
the fields of cryptology and signal processing for electronic warfare with
Thales and led UWB R\&D with Mitsubishi Electric from 2002 to 2007,
where he proposed two pioneering transceivers for short-range/high data
rates and large-range/low data rates, including telemetry. He developed multi-standard reconfigurable RF-IC at Renesas Design France until 2014. Since then, he is leading wireless activities for IRT b\raisebox{0.2mm}{\scalebox{0.7}{\textbf{$<>$}}}com.
\end{IEEEbiography}






\end{document}